\pdfoutput=1
\documentclass[aps,prl,twocolumn,superscriptaddress,amsmath,amssymb,longbibliography]{revtex4-2}
\usepackage[encapsulated]{CJK}
\usepackage{graphicx}
\usepackage[normalem]{ulem}
\usepackage{enumitem}
\usepackage[hidelinks]{hyperref}

\begin{document}
\begin{CJK*}{UTF8}{gbsn}
\title{Non-Hermitian edge accretion}
		\author{Wei Li (李维)}
		\affiliation{School of Physics and Electronics, Hunan University, Changsha 410082, China}

            \author{Ching Hua Lee}\email[Corresponding author: ]{phylch@nus.edu.sg}
	    \affiliation{Department of Physics, National University of Singapore, Singapore 117551, Singapore}

            \author{Fuxiang Li}\email[Corresponding author: ]{fuxiangli@hnu.edu.cn}
		\affiliation{School of Physics and Electronics, Hunan University, Changsha 410082, China}

		\begin{abstract}
        Under the non-Hermitian skin effect (NHSE), it is generally taken for granted that the open boundary condition (OBC) spectrum is encircled by the periodic boundary condition (PBC) spectrum. Intuitively, this is because boundaries tend to block the directed NHSE amplification, thereby suppressing $\operatorname{Im}(E)$.
        In this work, we analytically and numerically demonstrate how this conventional scenario can be dramatically violated:  In a phenomenon which we term non-Hermitian edge accretion (NHEA), \textit{strongly} coupled competing NHSE channels conspire to amplify a state whenever it bounces off an edge, instead of attenuating it. This unexpected behavior arises from underappreciated consequences of non-Bloch eigenspinor orthogonality, which control the reflection coefficients as well as the balance between propagating and evanescent waves. Operating in the strongly coupled regime, NHEA is fundamentally distinct from conventional critical NHSE scaling, which is characterized by amplification feedback generated in the \textit{bulk}, not the edges. Requiring only strongly coupled antagonistic NHSE chains with suitable mode orthogonality, the NHEA is platform-agnostic, and can be identified through periodically growing edge bounces in generic non-Hermitian metamaterials and quantum simulators.
		\end{abstract}
		\date{\today}
		\maketitle
\end{CJK*}

\textit{Introduction. }
Non-Hermitian systems possess complex energy spectra through gain or loss~\cite{Ashida2020,OkumaSato2023,topo1,lee2022exceptional,yang2026exceptional,sone2026hermitian}, as exhibited in photonic~\cite{El-Ganainy2018,Feng2017,Ozdemir2019,LiuEtAl2024Chiral,SunEtAl2024Floquet}, electrical~\cite{obc4,Ezawa2019Circuit,Tang2026NonAbelian,Wang2026ExceptionalFlat,Chen2026NonAbelianCircuit,Guo2024ScaleTailored,jiang2026engineering,rafi2022unconventional,hohmann2023observation,ZhangEtAl2024Impedance,zou2024experimental,li2019emergence} and acoustic metamaterials~\cite{Zhu2014,Fleury2015,ZhouEtAl2023Geometry,wang2026observation}, synthetic quantum platforms~\cite{Shen2025,Li2019,ZhaoEtAl2025Fermi,koh2026interacting} as well as bulk materials~\cite{li2024} and devices~\cite{OchkanEtAl2024}. If reciprocity is also broken, state propagation and amplification become direction-selective, famously known as the non-Hermitian skin effect (NHSE)~\cite{obc2,nh1,nh2,nh3,nh4,nh5,obc3,sensor1,pnhse,nh6,nh7,nh8,nh9,sfl4,KawabataNumasawaRyu2023,YoshidaEtAl2024Mott,QinLeeLi2025,qin2024kinked,xue2026topologically,hao2026hilbert,hao2026interacting,Liang2026Twist,Yuan2026Erratic,Jana2026NonMarkovian,Wang2026AcousticZ2,Zhang2026Harmonic,Zhang2025Algebraic,Hu2025ManyBody,Longhi2025Erratic,Kawabata2025Hopf,Wu2025Hybrid,MannaRoy2023Inner}. Crucially, a boundary interrupts such directed channels, reshaping long-time dynamics and the energy spectrum $E$~\cite{bcinhd,LiEtAl2024Dynamic,ShenChanLee2025,Franca2022ReflectedWaves,Xue2026EdgeDynamics}.
As such, we intuitively expect open boundary
conditions (OBCs) to suppress amplification~\footnote{For $\psi(t)=e^{-iEt}\psi(0)$, $\operatorname{Im}(E)$ is the amplitude growth rate, whereas $2\operatorname{Im}(E)$ is the growth rate of the norm.} relative to periodic boundary conditions (PBCs), i.e., $\max[\operatorname{Im}(E_{\rm OBC})]<\max[\operatorname{Im}(E_{\rm PBC})]$,  with the OBC spectrum hence lying
inside the PBC loop in the complex plane~\cite{onhse,nhse1,Kawabataprx1,Kawabataprx2,obc3}.

\begin{figure}
	\centering
	\includegraphics[width=\columnwidth]{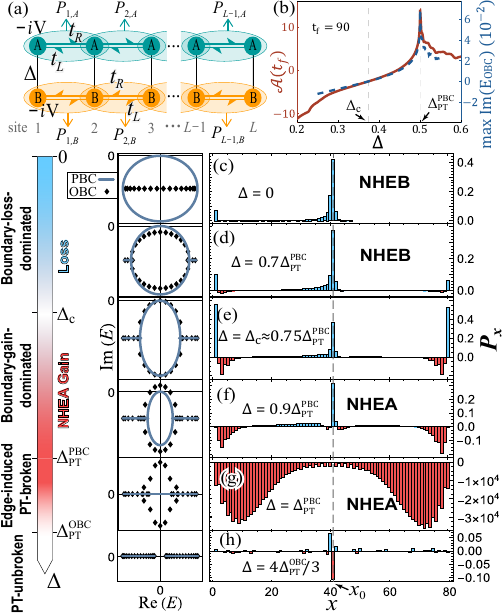}
	\caption{(a) The representative minimal NHEA model with antagonistic rightwards (A) and leftwards (B) Hatano-Nelson chains [Eq.~(\ref{hk})] strongly coupled by $\Delta$, with local dissipation $P_x=P_{x,A}+P_{x,B}$. (b) Dependence of NHEA extent $\mathcal A(t_f)$ [Eq.~(\ref{af})] (red solid, left axis) with $\Delta$, with the strongest accumulated gain (accretion) near $\Delta_{\rm PT}^{\rm PBC}$. It is closely proportional to  $\max[\operatorname{Im}(E_{\rm OBC})]$ (blue dashed curve, right axis).
    (c)--(h) Left Panel: Four distinct coupling regimes demarcated by thresholds $\Delta_c$,  $\Delta_{\rm PT}^{\rm PBC}$ and $\Delta_{\rm PT}^{\rm OBC}$ (see text for details). Middle Panel: PBC (gray lines) vs OBC (black diamonds) spectra. Right Panel: Corresponding $P_x$ for a walker initialized at $|x_0,B\rangle$ (blue/red denote loss/gain).
    (c-d) Ordinary NHSE regime $0\leq\Delta<\Delta_c$, with net loss (non-Hermitian edge burst, NHEB) when the walker encounters the left edge.
    (e) Balanced boundary-localized gain/loss (``perfectly reflective edges'') at the NHEA threshold $\Delta=\Delta_c$.
    (f) Strong NHEA regime marked by unconventional $\max[\operatorname{Im}(E_{\rm OBC})]>\max[\operatorname{Im}(E_{\rm PBC})]=0$ and significant edge accretion (red).
    (g) Peak NHEA at $\Delta=\Delta_{\rm PT}^{\rm PBC}$, with strongest edge accretion (red) and complex $E_{\rm OBC}$ with purely real $E_{\rm PBC}$ (opposite of usual NHSE).
    (h)  $\mathcal{PT}$-unbroken regime, where both OBC/PBC spectra are real and $P_x$ exhibits only small, featureless fluctuations.
 	Parameters used are $t_R=1$, $t_L=1/2$, $L=81$ and $x_0=41$. $P_x$ is integrated over the time window $[0,t_f]$, with $t_f\to\infty$ in (c)--(e) and $t_f=90$ otherwise. }
	\label{fig1}
\end{figure}

While the above intuition holds true in most cases in the literature~\cite{onhse,nhse1,Kawabataprx1,Kawabataprx2,obc3,Miri2019,Bender1998,Bender2007}, we find new interesting phenomenology when a boundary instead conspires to destabilize and \textit{amplify} the dynamics. Such scenarios, while conceptually exotic, are actually very simple to construct. We show that, just by \textit{strongly} coupling antagonistic NHSE chains, the OBC spectrum $E_{\rm OBC}$ may lie \textit{outside} the $E_{\rm PBC}$ loop, with $\max[\operatorname{Im}(E_{\rm OBC})]>\max[\operatorname{Im}(E_{\rm PBC})]$. We present one such archetypal example where $E_{\rm PBC}$ can become purely real while the finite-size OBC spectrum $E_{\rm OBC}$ can remain complex [Fig.~\ref{fig1}(g)], opposite of what happens under the usual NHSE~\cite{NBBT,PT,PT2,meng2025generalized,LeiLeeLi2024,PT3,PT4}.
This has profound dynamical implications: In a new phenomenon which we call non-Hermitian edge accretion (NHEA), a typical state amplifies and grows upon hitting a boundary, fundamentally distinct from usual NHSE.

To distinguish this NHEA behavior from the usual NHSE, we will quantify and compare their dynamical state evolution from a quantum walker perspective~\cite{PT3,PT4,obc5,obc7,qw2,qw3,qw4,qw5,qw6,LeeLonghi2020NonBloch}.
Suppose
$\max[\operatorname{Im}(E_{\rm PBC})]=0$, such that no exponential extinction or growth occurs away from the boundaries.
Under the usual NHSE, this gives rise to an ``edge burst'' when the surviving state is abruptly extinguished at a system boundary~\cite{eb1,eb2,eb3,eb4,eb5,eb6}. But under NHEA, we show that, paradoxically, the state grows in intensity---i.e., \textit{accretes}---upon encountering a boundary instead of being extinguished there.
As the state continues to evolve, it bounces around the system, growing by a characteristic NHEA-controlled extent at each bounce.

\textit{Physical setup for NHEA.--}
Far from being an elusive phenomenon, the NHEA can easily occur in strongly coupled NHSE chains. As the simplest archetypal example, consider coupling two Hatano--Nelson (HN) chains $h_A$ and $h_B$ with strength $\Delta$, as shown in Fig.~\ref{fig1}(a). For simplicity, we assume that $h_A(k)=h_B(-k)=t_L e^{ik}+t_R e^{-ik}$, with $t_R>t_L>0$, has equal and opposite nonreciprocities $t_L/t_R$ and $t_R/t_L$~\footnote{Arbitrary nonreciprocity ratios can be trivially obtained through a basis change~\cite{nh5,obc3}.}. The Bloch Hamiltonian is
\begin{equation}
	H(k)=
	\begin{pmatrix}
		h_A(k)-iV & \Delta\\
		\Delta & h_B(k)-iV
	\end{pmatrix},
	\label{hk}
\end{equation}
where $V=\mathrm{Re}\sqrt{(t_R-t_L)^2-\Delta^2}$ is the on-site loss, adjusted such that the PBC spectrum [gray loops in the left panels of Fig.~\ref{fig1}(c)--(h)]
\begin{equation}
	\begin{split}
		E_{\rm PBC}(k)=&-iV+(t_L+t_R)\cos k\\
		&\pm i\sqrt{(t_R-t_L)^2\sin^2k-\Delta^2}.
	\end{split}
	\label{epbc}
\end{equation}
satisfies $\max[\operatorname{Im}(E_{\rm PBC})]=0$ for all $\Delta$. This gives no possibility of exponential growth in the bulk~\cite{damp,deco}\footnote{For $0\leq\Delta<t_R-t_L$, the localized walker decays according to a universal power law $\sim t^{-1/2}$ [see Supplemental Material (SM), Sec.~V~\cite{SM}], agnostic to the NHSE strength.}, allowing us to clearly elucidate the novel bulk dynamics.

Importantly, we are concerned with the \textit{strong}-coupling regime $\Delta\sim\mathcal{O}(t_L),\mathcal{O}(t_R)$, characterized by strong feedback and interference that would lead to NHEA; this is completely distinct from the well-studied \textit{weak}-$\Delta$ ``critical NHSE'' regime~\cite{cnhse1,cnhse2,cnhse3,RafiUlIslamEtAl2022,LiuEtAl2020Critical,WangEtAl2025Coupled,qw2,ZhangEtAl2024Impedance,Percolation,XuEtAl2025Polariton}, where the chains effectively concatenate and exhibit anomalous dynamical scaling.

To unravel the NHEA dynamics, we consider a quantum walker initialized at site $x_0$ (away from edges $x=1,L$) on sublattice $B$ under OBCs, with $\langle x,s|\psi(0)\rangle=\delta_{x,x_0}\delta_{s,B}$. To keep track of the overall loss, we define the total norm $\mathcal{N}(t)\equiv\langle\psi(t)|\psi(t)\rangle$, which evolves according to
\begin{equation}
	\frac{\mathrm{d}}{\mathrm{d}t}\mathcal{N}(t)
	=i\langle\psi(t)|(H^\dagger-H)|\psi(t)\rangle.
	\label{dNdt}
\end{equation}
We also define the time-integrated local dissipation $P_x$ [see End Matter Eq.~(\ref{spx})], whose sum over the whole system satisfies $\sum_{x=1}^{L-1}P_x=\mathcal{N}(0)-\mathcal{N}(t_f)$. Positive and negative $P_x$ represent accumulated local loss and gain, respectively. To quantify the extent of accumulated gain from NHEA, irrespective of losses incurred by other mechanisms, we introduce the logarithmic accretion measure
\begin{equation}
 \mathcal A(t_f)\equiv\ln\left|\sum_{x=1}^{L-1}P_x(t_f)\Theta[-P_x(t_f)]\right|,
 \label{af}
\end{equation}
  where $\Theta$ is the Heaviside step function. The quantity inside the logarithm is the total magnitude of the accumulated negative dissipation, i.e., NHEA gain. At a common observation time $t_f=90$, Fig.~\ref{fig1}(b) shows that $\mathcal A(t_f)$ closely follows the trend of $\max[\operatorname{Im}(E_{\rm OBC})]$ across a broad regime where the interchain coupling $\Delta$ is strong and NHEA is pronounced, with both $\mathcal A(t_f)$ and $\max[\operatorname{Im}(E_{\rm OBC})]$ peaking near a key threshold which we will call $\Delta_{\rm PT}^{\rm PBC}=t_R-t_L$. Below, we provide a deep examination of how and why NHEA occurs.

\textit{Non-Hermitian edge accretion (NHEA) vs.\ edge burst: various regimes.--}
To resolve how the accretion develops, Fig.~\ref{fig1}(c)--(h) showcases the progression of regimes as the interchain coupling $\Delta$ increases from $0$. The left column shows the PBC (gray lines) and OBC (black diamonds) spectra, while the right column shows the time-integrated dissipation profiles $P_x$, for an initial excitation at $x=x_0$ at the center. Notably, as $\Delta$ increases, the OBC spectra expand significantly in $\operatorname{Im}(E)$ relative to the PBC spectra, leading to the four regimes indicated on the left, demarcated by critical values  $\Delta_c,\Delta_{\rm PT}^{\rm PBC}$ and $\Delta_{\rm PT}^{\rm OBC}$:
\begin{itemize}[leftmargin=0.3cm]
\item
\uline{At $\Delta=0$ [Fig.~\ref{fig1}(c)]}: In this uncoupled limit described by conventional NHSE, the OBC spectrum lies deep inside the PBC loop, with $\max[\operatorname{Im}(E_{\rm OBC})]<\max[\operatorname{Im}(E_{\rm PBC})]=0$. The local dissipation $P_x$ decreases from $x_0$ gradually via a bulk power law, leaving a pronounced blue lossy peak (``edge burst''~\cite{eb1,eb2,eb3,eb4,eb5,eb6}) at $x=1$, corresponding to rapid boundary dissipation, i.e., like a balloon bursting.

\item \uline{For $0<\Delta<\Delta_c$ [Fig.~\ref{fig1}(d)]}:
With nonvanishing interchain coupling $\Delta$, NHSE competition causes $\max[\operatorname{Im}(E_{\rm OBC})]$ to rise toward zero. The edge burst weakens; negative red $P_x$ pockets (gain) appear but the boundary loss still dominates.

\item \uline{At $\Delta=\Delta_c\equiv (t_R-t_L)\sqrt{t_R^2+t_L^2}/(t_R+t_L)$ [Fig.~\ref{fig1}(e)]}: This is the onset of NHEA, with $\max[\operatorname{Im}(E_{\rm OBC})]=\max[\operatorname{Im}(E_{\rm PBC})]=0$.
As such, the edges become ``perfectly reflective'' and edge gain (red) becomes significant, nearly canceling NHSE-type loss. This critical $\Delta_c$ threshold [see Eq.~(\ref{deltal}) below] does not represent a bulk transition; we will discuss the underlying boundary-scattering mechanism and norm-preserving condition later, with residual loss and system size scaling derived in SM Sec.~II~\cite{SM}.

\item \uline{For $\Delta_c<\Delta<\Delta_{\rm PT}^{\rm PBC}$ [Fig.~\ref{fig1}(f)]}:
In this deep NHEA regime, the OBC spectrum engulfs the PBC loop, with $\max[\operatorname{Im}(E_{\rm OBC})]>\max[\operatorname{Im}(E_{\rm PBC})]=0$, opposite of ordinary NHSE.
The negative boundary $P_x$ suggests \textit{boundary-gain-dominated} NHEA; exactly how this results from strongly coupling the two antagonistic chains will be discussed in the section below.
\item \uline{At $\Delta=\Delta_{\rm PT}^{\rm PBC}\equiv t_R-t_L$ [Fig.~\ref{fig1}(g)]}: This is when the NHEA is strongest, with the PBC spectrum entirely \textit{real} while finite-$L$ OBC eigenvalues remain complex. The strong accretion (red) extends from the edge deep into the bulk over the observation time.
Evidently the NHEA is thriving despite the absence of exponential growth in the $\mathcal{PT}$-symmetric bulk; details on the spectra, scaling trends, and the conditions for complex OBC branches are given in the SM, Secs. I and~VI~\cite{SM}.

\item \underline{For $\Delta_{\rm PT}^{\rm PBC}<\Delta\leq\Delta_{\rm PT}^{\rm OBC}\equiv t_R+t_L$}: As the interchain coupling $\Delta$ increases beyond $\Delta_{\rm PT}^{\rm PBC}$, the PBC bands are entirely real, yet the finite-chain OBC spectrum can contain complex-conjugate pairs, signaling edge-induced $\mathcal{PT}$ symmetry breaking. This arises from joint boundary matching of the four propagating modes in the band-overlap window (see SM, Secs.~III\,B and~VI~\cite{SM}). Finally at $\Delta = \Delta_{\rm PT}^{\rm OBC}$, both the OBC and PBC spectra become real, and the OBC state exhibits no exponential attenuation or amplification. The $P_x(t_f=90)$ profile for $\Delta=4\Delta_{\rm PT}^{\rm OBC}/3=2$ contains only small fluctuations [Fig.~\ref{fig1}(h)]---the expected featureless $\mathcal{PT}$-unbroken response.

\end{itemize}

    \begin{figure}[t]
	\centering
	\includegraphics[width=0.45\textwidth]{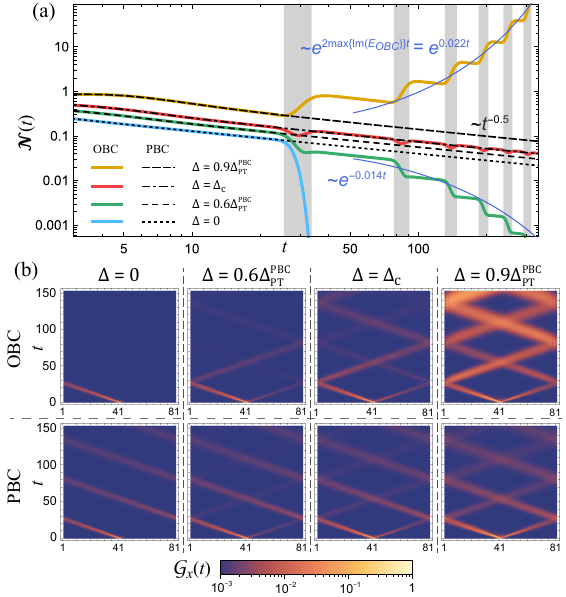}\\
	\caption{
        (a) Evolution of total norm $\mathcal{N}(t)$ of model Eq.~(\ref{hk}) under OBCs (solid) and PBCs (dashed). Under PBCs, $\mathcal{N}(t)$ follows a smooth $t^{-1/2}$ bulk envelope (dashed), whereas under OBCs this decay is punctuated by abrupt gain/loss epochs (gray windows) corresponding to boundary encounters. (b) Spatially resolved norm density $\mathcal{G}_x(t)\equiv\sum_{s=A,B}|\langle x,s|\psi(t)\rangle|^2$ under OBCs (top) and PBCs (bottom). In the isolated case ($\Delta=0$), the OBC wavepacket is extinguished at the left edge, whereas nonzero interchain coupling $\Delta$ generates increasingly strong counter-propagating wavepackets that grow and bounce off opposite edges and return; their arrivals coincide with the gray windows in (a), while the PBC trajectories remain uninterrupted. Across the coupling strengths shown, boundary encounters lead to attenuation ($\Delta=0.6\Delta_\text{PT}^\text{PBC}<\Delta_c$), perfect reflection ($\Delta=\Delta_c$), and NHEA amplification ($\Delta=0.9\Delta_\text{PT}^\text{PBC}>\Delta_c$), leading to an unconventional PBC-OBC spectral correspondence. Here $L=81$, $x_0=41$, $t_R=1$, $t_L=1/2$, $\Delta_{\rm PT}^{\rm PBC}=1/2$, and $\Delta_c=\sqrt{5}/6$.
    }\label{fig2}
    \end{figure}

\textit{NHEA from boundary encounters.--}
Having seen various regimes where the OBC spectrum can extend outside the PBC spectral locus and give rise to unconventional NHEA gain, we now trace its physical origin to boundary encounters by the evolved state. We examine the change in the state norm whenever it reflects off a boundary: as shown in Fig.~\ref{fig2}(a), under OBCs, the total norm $\mathcal{N}(t)$ exhibits abrupt gain or loss during boundary encounters, separated by smooth bulk-decay epochs, contrasting with monotonic PBC decay. Resolving the unit-cell norm density, $\mathcal{G}_x(t)=\sum_{s=A,B}|\langle x,s|\psi(t)\rangle|^2$, reveals the real-space dynamics [Fig.~\ref{fig2}(b)]: at nonzero interchain coupling $\Delta$, a localized wave packet splits into counterpropagating components that reflect off opposite boundaries, with attenuation or amplification depending on $\Delta$. Their arrival at the edges coincides with the abrupt changes in $\mathcal{N}(t)$, directly linking these changes to boundary reflection and scattering.

\begin{figure}[!t]
	\centering
	\includegraphics[width=0.45\textwidth]{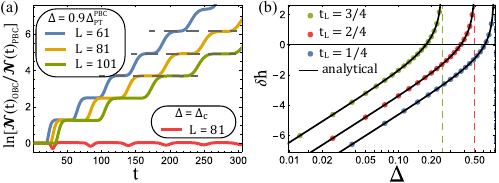}\\
	\caption{(a) Evolution of the logarithm of the ratio $\ln[\mathcal{N}_{\rm OBC}(t)/\mathcal{N}_{\rm PBC}(t)]$, which captures only the growth or attenuation from boundary encounters, being constant in the bulk.
   In the deep NHEA regime $\Delta=0.90\Delta_{\rm PT}^{\rm PBC}$ $(=0.45)$, each boundary encounter gives a significant $L$-independent step-up $\delta h>0$ representing gain;
   at the NHEA threshold $\Delta=\Delta_c=\sqrt{5}/6$ (pink), $\delta h\approx 0$ signifies perfect reflection. (b) Dynamically extracted logarithmic boundary step $\delta h$ (dots) as a function of $\Delta$, in excellent agreement with the analytic prediction (black curves) from Eq.~(\ref{dhanalytic}). $\delta h= 0$ crossings determine the NHEA thresholds $\Delta_c$, while $\delta h$ diverges logarithmically as
   $\Delta\to(\Delta_{\rm PT}^{\rm PBC})^-$. The dashed lines mark $\Delta_{\rm PT}^{\rm PBC}=t_R-t_L$: $0.25$ (green), $0.50$ (red), and $0.75$ (blue). Here $t_R=1$, with $x_0=(L+1)/2$ in (a) and $L=81$, $x_0=41$ in (b).}
	\label{fig3}
\end{figure}

For $0<\Delta<\Delta_{\rm PT}^{\rm PBC}$, we can isolate the boundary response by removing the common $t^{-1/2}$ bulk decay envelope (see SM Sec.~V\,A~\cite{SM}). This is done by taking the ratio of the OBC and PBC norms
\begin{equation}
h(t)\equiv\ln[\mathcal{N}_{\rm OBC}(t)/\mathcal{N}_{\rm PBC}(t)],
\label{ht}
\end{equation}
which exhibits flat plateaus during bulk propagation [Fig.~\ref{fig3}(a)]. The plateau widths are controlled by the velocities of the dominant modes at $k=\pm\pi/2$. For the upper-band dispersion $E_+(k)$, these velocities are  $v_{\leftarrow,\rightarrow}=\operatorname{Re}[\partial_k E_+(k)]_{k=\pm\pi/2}=\mp(t_R+t_L)\equiv\mp v$ [Fig.~\ref{fig2}(b)]. For midpoint initialization, the packets reach opposite boundaries at $t\approx(n+\tfrac12)T$ and return to the center at $t\approx(n+1)T$, where $T\equiv(L+1)/v$ is the duration for traversing across the system once ($n\in\mathbb{N}_0$). Sampling $h(t)$ at $t=nT$ thus cleanly keeps track of the effect of a simultaneous reflection event at both boundaries:
\begin{equation}
	\delta h\equiv h((n+1)T)-h(nT).
	\label{dhstep}
\end{equation}
From Fig.~\ref{fig3}(a), the plateau jumps $\delta h$ are all approximately the same height, independent of system length $L$, supporting their interpretation as isolated boundary reflection events. This suggests that each boundary encounter produces a characteristic norm multiplier, whose microscopic origin we examine next.

\textit{Physical anatomy of NHEA boundary encounters.--}
To uncover how boundary reflection can generate net NHEA amplification, we examine what is unusual in its boundary-scattering process. We expand an OBC eigenstate in generalized non-Bloch modes as $\psi_x=\sum_{j=1}^{4}c_j\beta_j^x\phi_j$, where $\beta=e^{ik}$ extends Bloch phase factors to complex wave numbers, $\beta_j$ are roots of $\det[H(\beta)-E_{\rm OBC}]=0$, and the spinors satisfy $H(\beta_j)\phi_j=E_{\rm OBC}\phi_j$. For fixed $0<\Delta<\Delta_{\rm PT}^{\rm PBC}$ and sufficiently large $L$, the boundary skin decay length is much smaller than $L$, allowing the two boundaries to be treated independently. Meanwhile, the OBC spectrum approaches the PBC spectrum~\cite{cnhse1,cnhse2,cnhse3}, whose maximal imaginary part is attained at $E=0$ under our choice of $V$; we therefore evaluate the leading local reflection at $E=0$.

The characteristic equation at $E_{\rm OBC}=0$ yields four roots: $(\beta_1,\beta_2,\beta_3,\beta_4)=(-i/\lambda,i,-i,i\lambda)$ with $\lambda>1$, as derived in Eq.~(\ref{eq:em-adelta}) in the End Matter. There are two counterpropagating bulk channels corresponding to the unit-modulus roots $\beta_2=i$ and $\beta_3=-i$, and we label them by spinors $\phi_L$ and $\phi_R$. There are also two evanescent skin modes corresponding to $\beta_1$ and $\beta_4$, localized with decay lengths $1/\ln\lambda$ at the left and right boundaries respectively. Because the propagating spinors are strictly orthogonal i.e. $\phi_L^\dagger\phi_R=0$, satisfying the open-boundary condition requires additional components from the skin modes. Crucially, these skin modes mediate reflection between the orthogonal propagating channels, a mechanism absent in the uncoupled NHSE limit. In more general models, such orthogonality in the propagating modes can also obstruct usual boundary condition matching and lead to violations in conventional NHSE behavior.

For our model, the boundary reflection is dominantly controlled by the angle $\theta\in(0,\pi/2)$ between the propagating and evanescent waves, with $\cos\theta=|\phi_L^\dagger\phi_1|=|\phi_R^\dagger\phi_4|$ [see End Matter and SM, Sec.~IV~\cite{SM}].
This is because $\cos\theta$ and $\sin \theta$ also represent the state amplitudes along the incident and outgoing channels, respectively. Canceling both channel components at the boundary therefore fixes the reflected-to-incident amplitude ratio to $\sin\theta/\cos\theta=\tan\theta$ in magnitude [End Matter Eqs.~(\ref{eq:em-BCleft}) and~(\ref{eq:em-BCright})].
With the phase conventions used there, the reflection amplitudes at the left and right boundaries are $r_L=-\tan\theta$ and $r_R=\tan\theta$, respectively [End Matter Eq.~(\ref{eq:em-reflection})].
Thus, upon simultaneous reflection at the two edges, each counterpropagating wavepacket acquires the norm factor $|r_L|^2=|r_R|^2=\tan^2\theta$.
This leads to a log relative rescaling factor [Eq.~(\ref{ht})]
\begin{equation}
	\begin{aligned}
		\delta h&\simeq\ln\tan^2\theta.
	\end{aligned}
	\label{dhscattering}
\end{equation}
Eq.~(\ref{dhscattering}) establishes the microscopic origin of NHEA: whether a boundary acts as an absorber ($\theta<\pi/4$) or an amplifier ($\theta>\pi/4$) is set, to leading order, by the geometric projection angle $\theta$ of the mediating skin spinor, which only exists due to the \textit{presence} of strongly competing NHSE modes afforded by strong antagonistic coupling.

The boundary step $\delta h$ in Eq.~(\ref{dhscattering}) can be analytically expressed in terms of the interchain coupling $\Delta$ through systematic boundary condition matching [see End Matter and SM, Sec.~IV~\cite{SM}]:
\begin{equation}
		\delta h(\Delta)
		\simeq-\ln\!\left[
		\frac{\cosh a-1}{\cosh a+1}
		\frac{\cosh b+\cosh a}{\cosh b-\cosh a}
		\right],
	\label{dhanalytic}
\end{equation}
where $a=\ln\lambda(\Delta)=\operatorname{arsinh}[(t_R+t_L)V/(2t_Rt_L)]$ and $b/2=\ln(\sqrt{t_R/t_L})$ are the inverse decay lengths of our evanescent skin mode and an isolated Hatano-Nelson chain respectively.
This analytic prediction matches numerical quantum-walk results with excellent  precision [Fig.~\ref{fig3}(b)]. It constitutes, to our knowledge, the first precise analytic result on a phenomenon that strongly violates usual generalized Brillouin zone (GBZ) predictions (for which only $|\beta_2|=|\beta_3|$ participates, and $\text{max}[\text{Im}(E_\text{OBC})]<\text{max}[\text{Im}(E_\text{PBC})]$).

From Eq.~(\ref{dhanalytic}) and Fig.~\ref{fig3}(b), for $\Delta<\Delta_{\rm PT}^{\rm PBC}$, increasing the interchain coupling $\Delta$ increases the accretion extent $\delta h$ by reducing $a$ and increasing $\theta$, converting boundary loss ($\delta h<0$) into gain ($\delta h>0$).
As $\Delta\to(\Delta_{\rm PT}^{\rm PBC})^-$ [dashed lines in Fig.~\ref{fig3}(b)], $\delta h$ undergoes a logarithmic divergence that can be traced to mode coalescence at a bulk exceptional point (EP) [see End Matter and SM, Sec.~IV~\cite{SM}].

The boundary amplification threshold $\Delta=\Delta_c$ [Fig. 1(e)], which marks the transition from net loss to net gain ($\delta h =0$), corresponds to the special value of $\theta=\pi/4$. At this threshold, the left and right skin spinors, $\phi_1=(\phi_L-\phi_R)/\sqrt{2}$ and $\phi_4=(\phi_R+\phi_L)/\sqrt{2}$, have equal-magnitude projections onto the incident and outgoing channels. Setting $\theta=\pi/4$ in Eq.~(\ref{dhanalytic}) gives norm-preserving reflection and vanishing $\delta h$, yielding
\begin{equation}
	\Delta_c=
	\frac{t_R-t_L}{t_R+t_L}
	\sqrt{t_R^2+t_L^2}.
	\label{deltal}
\end{equation}
Evidently, the onset of NHEA, marked by $\Delta_c$, is a qualitative transition in open-boundary wave matching of competing non-GBZ modes rather than a bulk Bloch spectral transition. Subtleties involving the variation of $\Delta_c$ with $L$ are derived in SM, Sec.~II~\cite{SM}.

\textit{Conclusion.--}
This work reveals that the physics behind antagonistic NHSE feedback lies far beyond the mere competition of skin amplification $|\beta_j|^x$. Specifically, the orthogonality of the non-Bloch eigenspinor modes can also substantially affect the response, in this case through non-Hermitian edge accretion (NHEA) where open boundaries serve to dramatically amplify rather than block skin amplification. This can cause the OBC spectrum to lie entirely \textit{outside} of the PBC spectrum, in stark contrast to usual NHSE where the PBC spectrum encircles the OBC spectrum. Operating in the \textit{strongly} coupled regime, it is conceptually distinct from usual critical NHSE, where weakly coupled NHSE chains experience PBC-like skin amplification feedback without boundary accretion.

 Together, these findings establish boundary-mediated channel conversion as an amplifying feedback mechanism: open boundaries can destabilize coupled nonreciprocal systems whose Bloch modes exhibit no exponential growth. Relying only on strong antagonistic NHSE coupling in a platform-agnostic manner, NHEA can manifest as boundary-induced amplification in non-Hermitian photonic~\cite{obc7,nh9}, acoustic~\cite{Zhang2026Harmonic,Fleury2015}, and electrical~\cite{Chen2026NonAbelianCircuit,Guo2024ScaleTailored,	jiang2026engineering,rafi2022unconventional,	hohmann2023observation,ZhangEtAl2024Impedance,zou2024experimental} metamaterials, as well as non-unitary quantum simulators~\cite{ZhaoEtAl2025Fermi,Li2019,shen2026simulating, 	chen2023high,Shen2025,koh2026interacting}.

\textit{Acknowledgements.}
This paper was supported by the National Natural Science Foundation of China (Grants No. 12675038), and the Fundamental Research Funds for the Central Universities of China. C.H.L. acknowledges support by the Singapore Ministry of Education (MOE) through the Academic Research Fund Tier-II Grant (MOE-T2EP50224-0007).

\bibliography{NHEA_References}

\clearpage
\section{End Matter}

\subsection{Definition of the signed accumulated local dissipation}\label{sec:em-local-dissipation}

\begin{figure}[t]
	\centering
	\includegraphics[width=0.9\columnwidth]{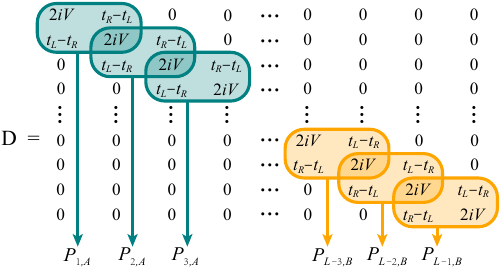}
	\caption{Structure of $D=H^\dagger-H$ in the basis ordered by chain ($A$ followed by $B$), with sites arranged in increasing $x$ within each chain. Green (yellow) blocks collect onsite and hopping terms on chain $A$ ($B$). Retaining only the $s=A$ or $s=B$ terms in Eqs.~\eqref{pon} and \eqref{pho}, with the onsite allocation in Eq.~\eqref{spx}, defines $P_{x,A}$ or $P_{x,B}$, respectively, such that $P_x=P_{x,A}+P_{x,B}$.}\label{figa2}
\end{figure}

The total norm $\mathcal N(t)\equiv\langle\psi(t)|\psi(t)\rangle$ obeys
\begin{equation}
	\frac{d\mathcal N}{dt}=i\langle\psi(t)|D|\psi(t)\rangle,
	\qquad D\equiv H^\dagger-H.
	\label{dt}
\end{equation}
For our coupled HN model [Eq.~\eqref{hk}], integrating $-d\mathcal N/dt$ over $[0,t_f]$ and grouping the diagonal and nearest-neighbor entries of $D$ (Fig.~\ref{figa2}) gives the position-resolved onsite and hopping dissipation contributions
\begin{align}
	P_{x,\rm onsite}(t_f)&=2V\sum_{s=A,B}\int_0^{t_f}dt\,|\langle x,s|\psi(t)\rangle|^2,\label{pon}\\
	P_{x,\rm hop}(t_f)&=2(t_R-t_L)\sum_{s=A,B}(\delta_{s,A}-\delta_{s,B})\nonumber\\
	&\quad\times\int_0^{t_f}dt\,\mathrm{Im}[\langle\psi(t)|x,s\rangle\langle x+1,s|\psi(t)\rangle].\label{pho}
\end{align}
While $P_{x,\rm onsite}\ge 0$ represents manifestly non-negative onsite loss, the current-like hopping dissipation $P_{x,\rm hop}$ can take either sign.
Sharing each interior onsite term equally between adjacent links, and each boundary term with its sole adjacent link, defines (every term on the right is evaluated at $t_f$)
\begin{equation}
	\begin{aligned}
		P_x(t_f)&=P_{x,\rm hop}\\
		&\quad+\begin{cases}
			P_{1,\rm onsite}+P_{2,\rm onsite}/2,&x=1,\\
			(P_{x,\rm onsite}+P_{x+1,\rm onsite})/2,&2\le x\le L-2,\\
			P_{L-1,\rm onsite}/2+P_{L,\rm onsite},&x=L-1
		\end{cases}
	\end{aligned}
	\label{spx}
\end{equation}
for $L\ge3$. Summing over links gives $\sum_{x=1}^{L-1}P_x(t_f)=\mathcal N(0)-\mathcal N(t_f)$: positive $P_x$ denotes accumulated loss and negative $P_x$ gain. We retain finite $t_f$ for growing dynamics; $t_f\to\infty$ is taken only when $\mathcal N(\infty)=0$.

\subsection{Mathematical origin of NHEA}
\label{sec:em-mathematical-origin}

To mathematically understand the NHEA, we first establish the boundary condition of the non-Bloch modes. We consider fixed $0<\Delta<\Delta_{\rm PT}^{\rm PBC}=t_R-t_L$, with $t_R>t_L>0$. Replacing $e^{ik}$ by $\beta$ in Eq.~\eqref{hk} gives four roots of $\det[H(\beta)-E]=0$ at any $E$. Expanding an OBC eigenstate as $\psi_x=\sum_jc_j\beta_j^x\phi_j$~\cite{NBBT,nh5}, we can parametrize $\phi_j\propto(\Delta,X_j)^T$, where $X_j=E+iV-t_L\beta_j-t_R\beta_j^{-1}$ and $X_{i,j}=X_i-X_j$. Imposing $\psi_0=\psi_{L+1}=0$ and eliminating $c_j$ gives the boundary quantization condition (see SM Sec.~III\,A~\cite{SM}):
\begin{equation}
	\begin{aligned}
		-X_{1,3}X_{2,4}(\beta_2\beta_4)^{L+1}
		+X_{1,2}X_{3,4}(\beta_3\beta_4)^{L+1}
		\simeq0,
	\end{aligned}
	\label{eq:em-dombeq}
\end{equation}
where the roots are ordered by modulus $|\beta_1|<|\beta_2|\le|\beta_3|<|\beta_4|$ and only the two leading terms are retained in the large-$L$ regime specified below.

To determine the long-time growth, we focus on eigenvalues with the largest imaginary part. As discussed in the main text, the plateau jumps $\delta h$ are virtually independent of system size $L$ [Fig.~\ref{fig3}(a)], motivating a large-$L$ expansion. In the thermodynamic limit, the OBC bulk spectrum follows the PBC bands, whose maximal imaginary part is fixed at zero by our choice $V=\sqrt{(t_R-t_L)^2-\Delta^2}$ and attained at $E=0$. There the roots are $\beta_{2,3}=\pm i$, $\beta_1=-i/\lambda$, and $\beta_4=i\lambda$, where the inverse decay length
\begin{equation}
	a\equiv\ln\lambda
	=\operatorname{arsinh}\!\left[
	\frac{(t_R+t_L)V}{2t_Rt_L}
	\right]
	\label{eq:em-adelta}
\end{equation}
decreases monotonically from $\ln(t_R/t_L)$ as $\Delta\to0^+$ to zero as $\Delta\to(t_R-t_L)^-$. For $L\gg1/a$, $|\beta_1|^{L+1}\ll 1\ll|\beta_4|^{L+1}$, which justifies Eq.~\eqref{eq:em-dombeq} near $E=0$.

Rearranging Eq.~\eqref{eq:em-dombeq} gives $(\beta_2/\beta_3)^{L+1}\simeq X_{1,2}X_{3,4}/(X_{1,3}X_{2,4})$. Evaluating this ratio near $E=0$ yields, to leading order, $(\beta_2/\beta_3)^{L+1}\simeq-\mathcal R^{-1}$, where
\begin{equation}
	\begin{aligned}
		\mathcal R
		&\equiv
		-\frac{X_{1,3}X_{2,4}}{X_{1,2}X_{3,4}}
		\bigg|_{E=0}\\
		&=
		\frac{\cosh a-1}{\cosh a+1}
		\frac{\cosh b+\cosh a}{\cosh b-\cosh a},
	\end{aligned}
	\label{eq:em-Rratio}
\end{equation}
with $b\equiv\ln(t_R/t_L)$. At finite $L$, the dominant OBC eigenvalue departs slightly from $E=0$. To retain the finite-size correction in the propagation factors, we write $\beta_2\simeq ie^{iq}$ and $\beta_3\simeq-ie^{-iq}$ with a small complex wave-number shift $q$. For even $L$, matching $(\beta_2/\beta_3)^{L+1}\simeq-\mathcal R^{-1}$ and choosing the solution nearest $q=0$ gives $q\simeq i\ln\mathcal R/[2(L+1)]$. Continuing Eq.~\eqref{epbc} to $\beta_2=e^{i(\pi/2+q)}$ gives $E_+(\beta_2)\simeq-(t_R+t_L)q=-vq$, yielding
\begin{equation}
	\max[\operatorname{Im}(E_{\rm OBC})]
	\simeq-\frac{t_R+t_L}{2(L+1)}\ln\mathcal R.
	\label{eq:em-quantization}
\end{equation}
For odd $L$, an additional real shift in $q$ leaves the leading imaginary term unchanged. Numerical results for $\max[\operatorname{Im}(E_{\rm OBC})]$ at odd $L$ agree with Eq.~\eqref{eq:em-quantization} (see SM Sec.~III\,A~\cite{SM}).

From Eqs.~\eqref{ht} and \eqref{dhstep}, $\delta h$ is the logarithmic OBC norm increment plus the PBC correction. For $\mathcal N_{\rm PBC}(t)\propto t^{-1/2}$, this correction is $\tfrac12\ln[(n+1)/n]$, which vanishes for $n\gg1$. Since the OBC norm grows asymptotically as
$e^{2t\max[\operatorname{Im}(E_{\rm OBC})]}$, we obtain $\delta h\simeq2T\max[\operatorname{Im}(E_{\rm OBC})]$. With $T=(L+1)/v$, Eq.~\eqref{eq:em-quantization} then yields $\delta h\simeq-\ln\mathcal R$, recovering the size-independent boundary step in Eq.~\eqref{dhanalytic}.

\subsection{Physical origin of NHEA}\label{sec:em-physical-origin}

The preceding subsection derives the gain-controlling factor $\mathcal R(\Delta)$ through boundary condition matching. To shed light on its physical meaning, consider local boundary scattering at $E=0$, the large-$L$ limit of the near-zero OBC branch (see also SM Sec.~IV~\cite{SM}). The dominant non-Bloch waves represented by $\beta_2=i$ and $\beta_3=-i$ carry left- and right-moving zero-growth spinors $\phi_L$ and $\phi_R$, with velocities $\mp v$, $v=t_R+t_L$ [Eq.~\eqref{epbc}]. Because $\phi_L^\dagger\phi_R=0$, these two-component spinors cannot cancel by themselves at an open edge: the evanescent roots $\beta_1=-i/\lambda$ and $\beta_4=i\lambda$ need to participate at the respective boundaries too.

To make these channel projections explicit, we perform the orthonormal basis rotation $\widetilde\psi=U_\Delta^\dagger\psi$ to map $\phi_R,\phi_L$ to $(1,0)^T,(0,1)^T$, respectively, using the matrix $U_\Delta=(\phi_R,\phi_L)$ given by
\begin{equation}
	U_\Delta=
	\begin{pmatrix}
		\sqrt{\dfrac{t_R-t_L+V}{2(t_R-t_L)}}&\sqrt{\dfrac{t_R-t_L-V}{2(t_R-t_L)}}\\
		-i\sqrt{\dfrac{t_R-t_L-V}{2(t_R-t_L)}}&i\sqrt{\dfrac{t_R-t_L+V}{2(t_R-t_L)}}
	\end{pmatrix}.
	\label{eq:em-UDelta}
\end{equation}
In this frame, the normalized skin spinors corresponding to $\beta_{1,4}$ take the forms
\begin{equation}
	\begin{gathered}
		\widetilde\phi_4=\begin{pmatrix}\cos\theta\\\sin\theta\end{pmatrix},\qquad
		\widetilde\phi_1=\begin{pmatrix}-\sin\theta\\\cos\theta\end{pmatrix},\\
		\cot^2\theta=\frac{|\phi_R^\dagger\phi_4|^2}{|\phi_L^\dagger\phi_4|^2}=\mathcal R,
	\end{gathered}
	\label{eq:em-skin-spinors}
\end{equation}
where the second equality follows from evaluating the spinor $\phi_4$ at $\beta_4=i\lambda$ [Eq.~\eqref{eq:em-Rratio}]. For $0<\theta<\pi/2$, $\cos\theta$ and $\sin\theta$ are the projection magnitudes along the incident and outgoing channels, respectively, at either wall. Explicitly, $\theta(\Delta)=\arctan[\mathcal R(\Delta)^{-1/2}]$, with the coupling dependence given by Eqs.~\eqref{eq:em-adelta} and \eqref{eq:em-Rratio}; increasing $\Delta$ therefore continuously rotates the skin spinor toward the outgoing channel.

To see how $\theta$ directly controls the extent of boundary reflection, write $\widetilde\psi_x=\sum_{j=1}^4\widetilde c_j\beta_j^x\widetilde\phi_j$. For $L\gg1/a$, we treat the two boundaries independently, omitting the exponentially small tails of the skin modes localized at the opposite boundaries. The local open-boundary conditions $\widetilde\psi_0=0$ and $\widetilde\psi_{L+1}=0$ become, respectively,
\begin{equation}
	\begin{aligned}
		\widetilde c_1
		\begin{pmatrix}-\sin\theta\\\cos\theta\end{pmatrix}
		+\widetilde c_2\begin{pmatrix}0\\1\end{pmatrix}
		+\widetilde c_3\begin{pmatrix}1\\0\end{pmatrix}=0,
	\end{aligned}
	\label{eq:em-BCleft}
\end{equation}
and
\begin{equation}
	\begin{aligned}
		\widetilde c_2\beta_2^{L+1}
		\begin{pmatrix}0\\1\end{pmatrix}
		+\widetilde c_3\beta_3^{L+1}
		\begin{pmatrix}1\\0\end{pmatrix}
		+\widetilde c_4\beta_4^{L+1}
		\begin{pmatrix}\cos\theta\\\sin\theta\end{pmatrix}=0.
	\end{aligned}
	\label{eq:em-BCright}
\end{equation}
Eliminating the skin amplitudes at both walls yields
\begin{equation}
	r_L\equiv\frac{\widetilde c_3}{\widetilde c_2}=-\tan\theta,\qquad
	r_R\equiv\frac{\widetilde c_2\beta_2^{L+1}}{\widetilde c_3\beta_3^{L+1}}=\tan\theta.
	\label{eq:em-reflection}
\end{equation}
Each boundary reflection thus multiplies the state norm by a factor
\begin{equation}
|r_L|^2=|r_R|^2=\tan^2\theta=\mathcal R^{-1}.
\end{equation}
This special relationship between boundary reflection and NHSE competition (through $\mathcal R$) underlines the origin of NHEA.
By tuning the boundary-matching angle $\theta$, the interchain coupling $\Delta$ directly dictates whether an open boundary attenuates ($\theta<\pi/4$) or amplifies ($\theta>\pi/4$) the incident wave.

For a midpoint initialized state, two separated wavepackets reach opposite walls once per interval $T=(L+1)/v$. Their norms add; both packets reflect once and acquire the factor $\tan^2\theta$.
As such, each boundary encounter gives the norm enhancement [Eqs.~\eqref{ht} and \eqref{dhstep}]
\begin{equation}
	\delta h\simeq\ln\tan^2\theta=-\ln\mathcal R
	\simeq2T\max[\operatorname{Im}(E_{\rm OBC})].
	\label{eq:em-step-link}
\end{equation}
This agrees with Eq.~\eqref{eq:em-quantization}. At $\theta=\pi/4$, these projections are equal and reflection preserves the norm to leading order -- this is none other than the ``perfectly reflective'' threshold $\Delta=\Delta_c$.

In the uncoupled limit $\Delta\to0^+$, $\theta\to0$ and the outgoing projection simply vanishes, recovering absorbing boundary behavior~\cite{cnhse2,cnhse3} without any NHEA.

In the strong NHEA limit $\Delta\to\Delta_{\rm PT}^{\rm PBC}=t_R-t_L$, $\beta_1\to\beta_3=-i$ and $\beta_4\to\beta_2=i$, while $\cos\theta\to0$; canceling an incident wave then requires an increasingly large skin amplitude. The rotated Hamiltonian $H_r(\beta)\equiv U_\Delta^\dagger H(\beta)U_\Delta$ explicitly expresses this as EP coalescence. Evaluated for the dominant channels,
\begin{equation}
	H_r(-i)=\begin{pmatrix}0&2i\Delta\\0&-2iV\end{pmatrix},\quad
	H_r(+i)=\begin{pmatrix}-2iV&0\\-2i\Delta&0\end{pmatrix}.
	\label{eq:em-Hrot}
\end{equation}
As $V\to0$ for $\Delta\to(\Delta_{\rm PT}^{\rm PBC})^-$, each matrix retains only one eigenvector, exposing the mode coalescence behind the positive divergence of $-\ln\mathcal R$ in Eq.~\eqref{dhanalytic}.

\end{document}


\begin{CJK*}{UTF8}{gbsn}
\begin{center}
{\large\bfseries Supplemental Material for ``Non-Hermitian edge accretion''\par}
\vspace{0.7em}
Wei Li (李维),\textsuperscript{1} Ching Hua Lee,\textsuperscript{2} and Fuxiang Li\textsuperscript{1}\par
\textsuperscript{1}School of Physics and Electronics, Hunan University, Changsha 410082, China\par
\textsuperscript{2}Department of Physics, National University of Singapore, Singapore 117551, Singapore
\end{center}
\end{CJK*}

    \renewcommand{\thefigure}{S\arabic{figure}}
    \renewcommand{\theequation}{S\arabic{equation}}
    \renewcommand{\thesection}{\Roman{section}}
    \setcounter{secnumdepth}{2}
    \renewcommand{\thesubsection}{\Alph{subsection}}
    This Supplemental Material presents the following calculations and derivations.

\tableofcontents

   \section{OBC spectra versus PBC spectra across different coupling strengths and system sizes}

\begin{figure}
	\centering
	\includegraphics[width=0.9\textwidth]{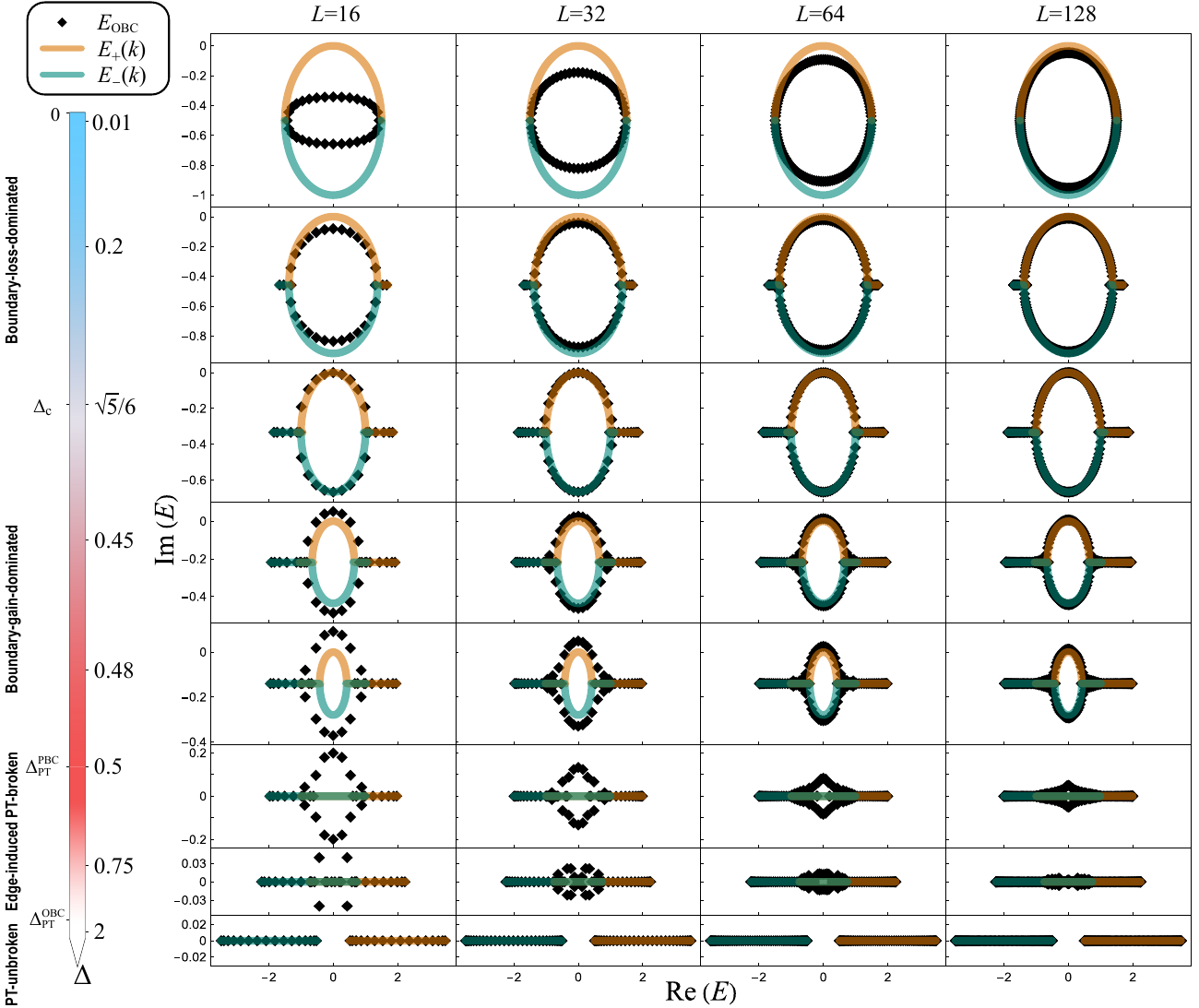}\\
	\caption{
			OBC and PBC spectra across coupling strengths and system sizes.
			Black diamonds denote OBC eigenvalues; orange and green curves denote the PBC bands $E_+(k)$ and $E_-(k)$, respectively.
			Columns correspond to $L=16,32,64$, and $128$, from left to right.
			Rows correspond to $\Delta=0.01$, $0.2$, $\sqrt{5}/6$, $0.45$, $0.48$, $0.5$, $0.75$, and $2$, from top to bottom.
			Here $t_R=1$, $t_L=1/2$, and $V=\mathrm{Re}\sqrt{(t_R-t_L)^2-\Delta^2}$.
			The color strip on the left identifies the four coupling regimes and marks the characteristic thresholds $\Delta_c=\sqrt{5}/6\approx0.373$, $\Delta_{\rm PT}^{\rm PBC}=0.5$, and $\Delta_{\rm PT}^{\rm OBC}=1.5$.
			As $\Delta$ increases, the PBC loop contracts into two partially overlapping real bands, which subsequently move apart.
			The upper OBC spectral edge crosses from negative to positive imaginary energies near $\Delta_c$, and complex OBC pairs remain visible at $\Delta=0.5$ and $0.75$ despite the entirely real PBC spectrum.
			At $\Delta=2$, both spectra are real.
			Increasing $L$ reduces the imaginary deviations of the OBC spectrum from the PBC bands.
			Vertical-axis scales differ between rows.
		}
		\label{spectra}
	\end{figure}

   The main text illustrates the four dynamical regimes of NHEA at a fixed system size ($L=81$). Figure~\ref{spectra} traces their spectral evolution over a range of interchain couplings $\Delta$ for $L=16,32,64$, and $128$, showing how the OBC--PBC comparison evolves with system size. We use the same uniform loss $V=\mathrm{Re}\sqrt{(t_R-t_L)^2-\Delta^2}$ as in the main text. For $\Delta<\Delta_{\rm PT}^{\rm PBC}$, this loss shifts the upper PBC spectral edge to zero imaginary energy; for $\Delta\geq\Delta_{\rm PT}^{\rm PBC}$, $V=0$ and the PBC bands are entirely real. Reading down each column reveals the evolution through the four coupling regimes, while comparing columns shows how the spectral differences between OBCs and PBCs depend on system size.

   For $0<\Delta<\Delta_{\rm PT}^{\rm PBC}$, increasing $\Delta$ progressively contracts the PBC loop, while horizontal segments extend outward from it at $\mathrm{Im}(E)=-V$. The rows at $\Delta=0.45$ and $0.48$ resolve the shrinking loop as it approaches the real axis. At $\Delta=\Delta_{\rm PT}^{\rm PBC}=0.5$, the loop collapses, leaving two partially overlapping line segments on the real axis. Above this threshold, the two real PBC bands move apart as $\Delta$ increases, reducing their overlap. The overlap vanishes at $\Delta=\Delta_{\rm PT}^{\rm OBC}=1.5$, and a band gap opens at larger couplings, as illustrated by the $\Delta=2$ row.

   Against this evolving PBC spectrum, the OBC eigenvalues reveal the change from boundary attenuation to amplification. At $\Delta=0.01$ and $0.2$, the upper OBC spectral edge lies below zero, corresponding to the boundary-loss-dominated regime. This negative offset decreases as $\Delta$ approaches $\Delta_c=\sqrt{5}/6$, where the upper OBC and PBC edges nearly coincide, corresponding to norm-preserving boundary reflection to leading order. Above $\Delta_c$, the upper OBC edge extends into $\mathrm{Im}(E)>0$, signaling boundary amplification. At $\Delta=0.45$ and $0.48$, the OBC spectrum therefore extends beyond the contracting PBC loop in the imaginary direction. At $\Delta=0.5$, complex-conjugate OBC pairs remain even though both PBC bands are entirely real, demonstrating edge-induced $\mathcal{PT}$ breaking. Increasing $\Delta$ further reduces their imaginary extent, as illustrated at $\Delta=0.75$. Finite-size boundary quantization determines where these pairs return to the real axis, allowing a fully real OBC spectrum to emerge below $\Delta_{\rm PT}^{\rm OBC}$ for a given $L$. For $\Delta\geq\Delta_{\rm PT}^{\rm OBC}$, both spectra are real for any $L$, as illustrated by the $\Delta=2$ row.

   Across each row, increasing $L$ makes the OBC spectrum denser and brings it closer to the PBC bands. Both the negative offset of the upper OBC edge below $\Delta_c$ and its positive excursion above $\Delta_c$ decrease with size. For fixed $0<\Delta<\Delta_{\rm PT}^{\rm PBC}$ away from the endpoints, each boundary reflection imparts an $L$-independent gain or loss factor to leading order, while the travel time between successive boundary encounters grows with $L$. The resulting amplification or attenuation rate therefore decreases as $L^{-1}$, accounting for the approach of the upper OBC spectral edge to its PBC counterpart, as derived in Sec.~\ref{sec:approxime}. Thus, the coupling-induced changes in the imaginary spectral extent are more pronounced for smaller systems, while their overall amplitude decreases with increasing $L$. In the real-PBC regime, boundary quantization produces oscillatory variations with $\Delta$ beneath an upper envelope that decreases with $L$. Moreover, at $\Delta=\Delta_{\rm PT}^{\rm PBC}$, the shrinking imaginary extent coexists with an extensive number of complex OBC eigenvalues in the large-$L$ estimate of Sec.~\ref{sec:complexnum}.

\section{Derivation of the threshold coupling \texorpdfstring{$\Delta_c$}{Delta(c)} for perfectly reflective edges and OBC gain}\label{sec:threshold}

In this section, we derive the analytical expression for the boundary amplification threshold $\Delta_c$ from the open-boundary conditions. As illustrated in Fig.~\ref{spectra}, this $L$-independent coupling separates the boundary-loss-dominated and boundary-gain-dominated regimes, with $\max[\operatorname{Im}(E_{\rm OBC})]=\max[\operatorname{Im}(E_{\rm PBC})]=0$ to leading scale-free order. The edges therefore preserve the reflected norm to leading order. We then determine the finite-$L$ residual loss and its size scaling: it is exponentially small for even $L$, whereas for odd $L$ the standing-wave phase shift produces a leading algebraic correction proportional to $(L+1)^{-2}$.

We take $t_R>t_L>0$ and consider $0<\Delta<\Delta_{\rm PT}^{\rm PBC}=t_R-t_L$, with the tuned loss $V=\mathrm{Re}\sqrt{(t_R-t_L)^2-\Delta^2}$ fixing the zero-growth bulk momenta at $k=\pm\pi/2$. Evaluating the boundary equation at $E=0$ first identifies $\Delta_c$ through cancellation of its dominant large-$L$ contribution. Retaining the finite-size corrections then resolves the remaining loss for even and odd $L$. The real-PBC regime $\Delta\geq\Delta_{\rm PT}^{\rm PBC}$ is treated separately in Secs.~\ref{sec:strongcoupl} and~\ref{sec:complexnum}.

We start from the standard non-Bloch (GBZ) ansatz~\cite{NBBT,nh5,meng2025generalized}
\begin{equation}
	\psi_x=\sum_{j=1}^{4}c_j\beta_j^x\phi(\beta_j),
	\label{sgbz}
\end{equation}
with the characteristic equation
\begin{equation}
	{\rm det}[H(\beta)-E_{\rm OBC}]=0,
	\label{sHbeta}
\end{equation}
where $H(\beta)$ is obtained by generalizing the Bloch phase factor $e^{ik}$ to complex $\beta$ in $H(k)$.
The real-space Schr\"{o}dinger equation of the coupled HN model reads $t_R\psi_{x-1}^{\rm A}+\Delta\psi_{x}^{\rm B}+t_L\psi_{x+1}^{\rm A}-iV\psi_{x}^{\rm A}=E_{\rm OBC}\psi_{x}^{\rm A}$ and $t_L\psi_{x-1}^{\rm B}+\Delta\psi_{x}^{\rm A}+t_R\psi_{x+1}^{\rm B}-iV\psi_{x}^{\rm B}=E_{\rm OBC}\psi_{x}^{\rm B}$. Imposing open boundary conditions $\psi_0^{s} = \psi_{L+1}^{s} = 0$, we obtain
\begin{equation}
	\begin{cases}
		\Delta\psi_{1}^{\rm B}+t_L\psi_{2}^{\rm A}-iV\psi_{1}^{\rm A}=E_{\rm OBC}\psi_{1}^{\rm A} \\
		\Delta\psi_{1}^{\rm A}+t_R\psi_{2}^{\rm B}-iV\psi_{1}^{\rm B}=E_{\rm OBC}\psi_{1}^{\rm B} \\
		t_R\psi_{L-1}^{\rm A}+\Delta\psi_{L}^{\rm B}-iV\psi_{L}^{\rm A}=E_{\rm OBC}\psi_{L}^{\rm A}\\
		t_L\psi_{L-1}^{\rm B}+\Delta\psi_{L}^{\rm A}-iV\psi_{L}^{\rm B}=E_{\rm OBC}\psi_{L}^{\rm B}
	\end{cases}
	\label{obc}
\end{equation}
Substituting the ansatz Eq.~(\ref{sgbz}) into the boundary equation Eq.~(\ref{obc}) yields the condition for nontrivial coefficients,
\begin{equation}
	\begin{vmatrix}
		1 & 1 & 1 & 1\\
		X_1 & X_2 & X_3 & X_4 \\
		(\beta_1)^{L+1} & (\beta_2)^{L+1} & (\beta_3)^{L+1} & (\beta_4)^{L+1} \\
		X_1(\beta_1)^{L+1} & X_2(\beta_2)^{L+1} & X_3(\beta_3)^{L+1} & X_4(\beta_4)^{L+1}
	\end{vmatrix}=0.
	\label{det}
\end{equation}
Here we choose $\phi(\beta_j)=(\Delta,X_j)^T$, with $X_j=E_{\rm OBC}+iV-t_L\beta_j-t_R\beta_j^{-1}$; the common factors of $\Delta$ have been removed from the determinant. Eq.~(\ref{det}) can then be rewritten as
\begin{equation}
	X_{1,4}X_{2,3}[(\beta_1\beta_4)^{L+1}+(\beta_2\beta_3)^{L+1}]-X_{1,3}X_{2,4}[(\beta_1\beta_3)^{L+1}+(\beta_2\beta_4)^{L+1}]+X_{1,2}X_{3,4}[(\beta_1\beta_2)^{L+1}+(\beta_3\beta_4)^{L+1}]=0,
	\label{sobc}
\end{equation}
where $X_{i,j}=t_L(\beta_j-\beta_i)+t_R(\beta_j^{-1}-\beta_i^{-1})\ (i,j=1,...,4)$. The factors $(\beta_i\beta_j)^{L+1}$ make the boundary quantization condition depend on whether $L$ is even or odd. We therefore treat the two cases separately. Up to Eq.~(\ref{sobc}), we have applied the standard non-Bloch boundary-matching framework. Our contribution is to use this boundary equation to derive the analytical threshold $\Delta_c$ separating boundary loss from gain, and to determine the parity-dependent finite-size residual loss at this threshold.

\subsection{Even \texorpdfstring{$L$}{L}: threshold and finite-size correction}

At the threshold, the leading size-dependent contribution to the boundary equation cancels. Evaluating Eq.~(\ref{sobc}) at zero energy identifies this coupling, and retaining the remaining boundary mismatch gives the finite-size displacement of the OBC eigenvalue.

\begin{figure}[t]
	\centering
	\includegraphics[width=0.85\textwidth]{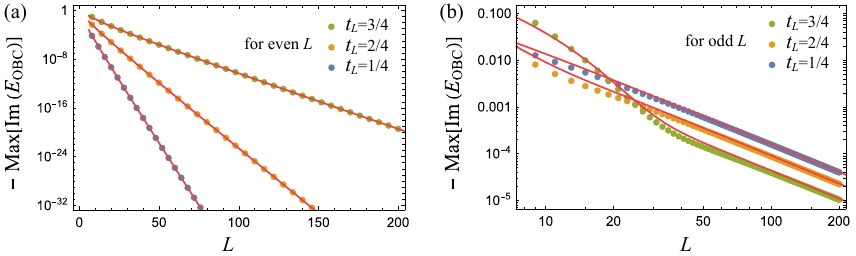}\\
	\caption{The negative of the largest imaginary OBC eigenvalue $-\max[\mathrm{Im}(E_{\rm OBC})]$ at the threshold $\Delta=\Delta_c$ for several $t_L$, with the tuned loss $V=V_c$, as a function of system size. (a) Even $L$: symbols are from numerical diagonalization, red lines the analytical prediction Eq.~(\ref{seveneven}) (exponentially small in $L$). (b) Odd $L$: symbols  are from numerical diagonalization, red lines Eq.~(\ref{soddthres}) (a crossover estimate). In both panels, $t_R=1$. The green points ($t_L=3/4$) show a pronounced crossover: Eq.~(\ref{soddthres}) estimates an exponential contribution $\lambda_c^{-L-1}/(L+1)$ at smaller sizes, where it dominates, and inverse-square decay $(L+1)^{-2}$ at large $L$.}
	\label{ime}
\end{figure}

For the even sizes shown in Fig.~\ref{spectra}, numerical diagonalization places the OBC eigenvalue with the largest imaginary part on the imaginary axis near the origin. We provisionally set $E_{\rm OBC}=0$ to identify the candidate threshold $\Delta_c$. The four roots of $\det[H(\beta)-0]=0$ are $\beta_1=-i\lambda_c^{-1}$, $\beta_4=i\lambda_c$, $\beta_2=i$, $\beta_3=-i$, with
\begin{equation}
	\lambda_c=\frac{(t_R+t_L)\sqrt{(t_R-t_L)^2-\Delta_c^2}+\sqrt{(t_R^2+t_L^2)^2-(t_R+t_L)^2\Delta_c^2}}{2t_Rt_L}.
	\label{lambdac}
\end{equation}
Substituting into Eq.~(\ref{sobc}) reduces the boundary equation to
\begin{equation}
	 2X_{1,4}X_{2,3}+(X_{1,3}X_{2,4}+X_{1,2}X_{3,4})(\lambda_c^{L+1}+\lambda_c^{-L-1})=0.
	\label{sobc2}
\end{equation}
For $L\ln\lambda_c\gg1$, the skin tails from opposite boundaries have negligible overlap, and the factor $\lambda_c^{L+1}+\lambda_c^{-L-1}$ dominates Eq.~(\ref{sobc2}). The threshold is fixed by cancellation of its coefficient: $X_{1,3}X_{2,4}+X_{1,2}X_{3,4}=0$. Solving this condition for $\Delta_c$ gives
\begin{equation}
	\Delta_c=\frac{t_R-t_L}{t_R+t_L}\sqrt{t_R^2+t_L^2}.
	\label{sobc4}
\end{equation}
This even-$L$ calculation yields the same $L$-independent boundary threshold as the norm-preserving reflection condition in the End Matter; parity affects the residual finite-size loss. After the $L$-dependent coefficient has fixed the unique candidate $\Delta_c$, the term $2X_{1,4}X_{2,3}$ in Eq.~(\ref{sobc2}) remains nonzero. Thus $E_{\rm OBC}=0$ is not an exact OBC eigenvalue at finite $L$ even at the threshold $\Delta_c$. We therefore look for the OBC eigenvalue with the largest imaginary part slightly displaced from the origin along the imaginary axis, $E_{\rm OBC}=i\epsilon$ with $|\epsilon|\ll 1$. The leading-order boundary mismatch is controlled by the unit-modulus roots $\beta_{2,3}=\pm i$, whose small phase displacement is related linearly to $E_{\rm OBC}$ through the local dispersion. The off-unit-circle roots $\beta_{1,4}$ shift only at subleading order in the boundary equation and are kept at their zero-energy values. Solving $\det[H(\beta)-i\epsilon]=0$ to leading order then gives

\begin{equation}
	\beta_1\approx -i\lambda_c^{-1},\qquad
	\beta_4\approx i\lambda_c,\qquad
	\beta_2\approx ie^{iq_i},\qquad
	\beta_3\approx -ie^{-iq_i},
	\label{betaeps}
\end{equation}
where $q_i=-i\epsilon/(t_R+t_L)$. Substituting Eq.~(\ref{betaeps}) into Eq.~(\ref{sobc}) and dropping the subleading $\lambda_c^{-L-1}$ contributions, we obtain
\begin{equation}
	2X_{1,4}X_{2,3}+X_{1,3}X_{2,4}\lambda_c^{L+1}e^{iq_i(L+1)}+X_{1,2}X_{3,4}\lambda_c^{L+1}e^{-iq_i(L+1)}\approx0.
	\label{sqeven0}
\end{equation}
Combining Eqs.~(\ref{lambdac}), (\ref{sobc4}), and (\ref{betaeps}), we obtain $X_{1,4}X_{2,3}\approx2X_{1,3}X_{2,4}$ and $X_{1,3}X_{2,4}+X_{1,2}X_{3,4}\approx0$. Eq.~(\ref{sqeven0}) can therefore be reduced to
\begin{equation}
	\sin[q_i(L+1)]\approx2i\lambda_c^{-L-1}.
	\label{sqeven1}
\end{equation}
By applying the small-angle approximation $\sin[q_i(L+1)]\approx q_i(L+1)$, we obtain
\begin{equation}
	q_i\approx \frac{2i}{L+1}\lambda_c^{-L-1}.
	\label{sqeven}
\end{equation}
Accordingly, we obtain $E_{\rm OBC}=i\epsilon=-(t_R+t_L)q_i$, namely,
\begin{equation}
	\max[\mathrm{Im}(E_{\rm OBC})]\approx -\frac{2(t_R+t_L)}{L+1}\lambda_c^{-L-1},
	\qquad L\ {\rm is \  even}
	\label{seveneven}
\end{equation}
where $\lambda_c$ is given by Eq.~(\ref{lambdac}).
Hence, the even-$L$ correction is exponentially small, and our numerical results closely match Eq.~(\ref{seveneven}), as shown in Fig.~\ref{ime}(a).

\subsection{Odd \texorpdfstring{$L$}{L}: finite-size correction imposed by the OBC standing-wave quantization}

For odd $L$, the boundary conditions impose a different momentum quantization from the even-$L$ case. Assuming, at this stage, that the OBC eigenvalue with the largest imaginary part lies on the PBC band, the real shift $q_r$ measures how far this quantization moves the allowed momentum away from the zero-growth point $k=\pi/2$ along that band. To determine $q_r$, we write
\begin{equation}
	\beta_1\approx -i\lambda_c^{-1},\qquad
	\beta_4\approx i\lambda_c,\qquad
	\beta_2\approx ie^{iq_r},\qquad
	\beta_3\approx -ie^{-iq_r}.
	\label{soddbetaeps}
\end{equation}
Substituting Eq.~(\ref{soddbetaeps}) into Eq.~(\ref{sobc}), evaluating $X_{i,j}$ at their zero-energy values, and dropping terms proportional to $\lambda_c^{-L-1}$ gives
\begin{equation}
	2X_{1,4}X_{2,3}-X_{1,3}X_{2,4}\lambda_c^{L+1}e^{iq_r(L+1)}+X_{1,2}X_{3,4}\lambda_c^{L+1}e^{-iq_r(L+1)}\approx0.
	\label{sqodd0}
\end{equation}
As in the even-$L$ case, at $\Delta=\Delta_c$ we have $X_{1,4}X_{2,3}\approx2X_{1,3}X_{2,4}$ and $X_{1,3}X_{2,4}+X_{1,2}X_{3,4}\approx0$. Eq.~(\ref{sqodd0}) can therefore be reduced to
\begin{equation}
	\lambda_c^{L+1}e^{iq_r(L+1)}+\lambda_c^{L+1}e^{-iq_r(L+1)}=2\lambda_c^{L+1}\cos[q_r(L+1)]\approx4,
	\label{sqodd1}
\end{equation}
which yields
\begin{equation}
 q_r\approx\frac{1}{L+1}\arccos(2\lambda_c^{-L-1})\approx\frac{\pi}{2(L+1)}.
\end{equation}
Thus, even when the near-zero OBC energies are approximated by points on the PBC band, odd-$L$ quantization prevents the eigenvalue with the largest imaginary part from reaching $E=0$.

The purely imaginary shift $q_i$ describes the additional displacement away from the PBC band. We retain the even-$L$ expression for $q_i$ [Eq.~(\ref{sqeven})] as a phenomenological contribution. Combining the two shifts as $\beta_2\approx i e^{i(q_r+q_i)}$ and expanding the dispersion around $(\beta,E)=(i,0)$ gives
\begin{equation}
	E_{\rm OBC}\approx -(t_R+t_L)(q_i+q_r)-i\frac{(t_R-t_L)^2}{2V_c}(q_i+q_r)^2.
	\label{soddexp}
\end{equation}
This gives the following crossover estimate for the odd-$L$ finite-size correction:
\begin{equation}
	\max[\mathrm{Im}(E_{\rm OBC})]
	\approx
	-\frac{2(t_R+t_L)}{L+1}\lambda_c^{-L-1}-\frac{(t_R-t_L)^2\pi^2}{8V_c(L+1)^2},
	\qquad L\ {\rm is \  odd},
	\label{soddthres}
\end{equation}
where $\lambda_c$ is given by Eq.~(\ref{lambdac}) and
\begin{equation}
	V_c\equiv V|_{\Delta=\Delta_c}=\frac{(t_R-t_L)\sqrt{2t_Rt_L}}{t_R+t_L}.
	\label{vc}
\end{equation}
The algebraic term follows from the odd-$L$ momentum quantization.
To estimate the crossover at smaller sizes, we supplement it with
the exponentially small contribution obtained for even $L$.
The resulting expression is a phenomenological crossover estimate,
compared with numerical diagonalization in Fig.~\ref{ime}(b).

In summary, $\Delta_c$ marks the coupling at which $\max[\mathrm{Im}(E_{\rm OBC})]$ is maximally insensitive to system size. For finite $L$ the largest imaginary OBC eigenvalue is not exactly zero, but its deviation is exponentially small for even $L$ and algebraically $\propto(L+1)^{-2}$ for odd $L$; both corrections vanish rapidly with increasing $L$. For the system sizes relevant to the quantum-walk dynamics in the main text (e.g.\ $L=81$), the residual finite-size effects are too small to affect the algebraic decay within a finite observation window. The value of $\Delta_c$ itself is exact in the scale-free sense. The convergence of the finite-size gain threshold to $\Delta_c$ is analyzed in Sec.~\ref{sec:exactthreshold}.

\section{Approximate expressions for \texorpdfstring{$\max[\mathrm{Im}(E_{\rm OBC})]$}{max[Im(E(OBC))]} in different coupling regimes}\label{sec:approxime}

We use the exact OBC boundary equation~\eqref{sobc} to derive large-$L$ approximations for $\max[\mathrm{Im}(E_{\rm OBC})]$. These expressions identify the change from boundary loss to gain and its finite-size scaling.

We present different derivations for the complex-PBC and real-PBC intervals, demarcated by $\Delta_{\rm PT}^{\rm PBC}=t_R-t_L$. As in Sec.~\ref{sec:threshold}, the strategy is to first identify the root configuration near zero energy and then extract its finite-size shift from the boundary equation. The first subsection treats the regime $0<\Delta<t_R-t_L$, where the PBC spectrum is complex and touches the origin at $k=\pm\pi/2$. It also shows how the finite-size OBC gain threshold converges to the $L$-independent boundary amplification threshold $\Delta_c$. The second subsection addresses the edge-induced $\mathcal{PT}$-broken regime, $t_R-t_L<\Delta<t_R+t_L$, where the PBC spectrum is purely real and boundary quantization determines whether a given finite chain has complex OBC branches.

\subsection{Boundary-loss-dominated and boundary-gain-dominated regimes: \texorpdfstring{$0<\Delta<\Delta_{\rm PT}^{\rm PBC}$}{0 < Delta < Delta(PT, PBC)}}

Throughout these two regimes, $0<\Delta<t_R-t_L$, $V=\mathrm{Re}\sqrt{(t_R-t_L)^2-\Delta^2}$ and the PBC spectrum touches $E=0$ at $k=\pm\pi/2$ (equivalently $\beta=\pm i$).
Numerical diagonalization shows that the OBC eigenvalue with the largest imaginary part approaches this zero-energy point as $L$ increases at fixed coupling (Fig.~\ref{spectra}). Thus, as in Sec.~\ref{sec:threshold}, the relevant finite-size correction can be extracted by expanding the OBC boundary equation around the zero-energy root configuration.

At $E_{\rm OBC}=0$, the characteristic equation ${\rm det}[H(\beta)-E_{\rm OBC}]=0$ has the two exact unit-modulus roots $\beta_2=i$ and $\beta_3=-i$. By solving the characteristic equation Eq.~(\ref{sHbeta}), the remaining pair are $\beta_1=-i/\lambda$ and $\beta_4=i\lambda$ with
\begin{equation}
	\lambda=\frac{(t_R+t_L)V+\sqrt{(t_R^2+t_L^2)^2-(t_R+t_L)^2\Delta^2}}{2t_Rt_L}>1.
	\label{slambdagen}
\end{equation}

For finite $L$, the boundary condition is generally not satisfied exactly at $E_{\rm OBC}=0$. The OBC eigenvalue closest to the origin is therefore obtained by shifting the unit-modulus pair slightly away from $\beta=\pm i$, while keeping the off-unit-circle pair at its leading zero-energy value, i.e. the approximations:
\begin{equation}
	\beta_1\approx -i\lambda^{-1},\qquad
	\beta_4\approx i\lambda,\qquad
	\beta_2\approx ie^{iq},\qquad
	\beta_3\approx -ie^{-iq}.
	\label{sbetashiftgen}
\end{equation}
At leading order, odd $L$ adds a real standing-wave phase shift to the even-$L$ quantization (Sec.~\ref{sec:threshold}). We therefore use even $L$ to derive the leading imaginary shift, which is common to both parities.

\begin{figure}[t]
	\centering
	\includegraphics[width=0.96\textwidth]{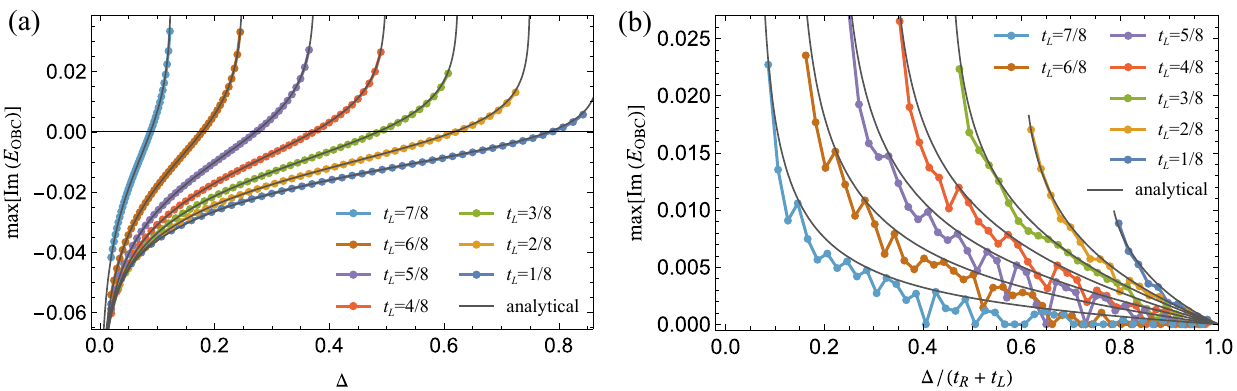}\\
	\caption{Numerical and analytical results for the largest imaginary OBC eigenvalue $\max[\mathrm{Im}(E_{\rm OBC})]$ at $t_R=1$ and $L=81$ for several $t_L$. Symbols joined by colored lines are numerical diagonalization results; gray curves are the analytical approximations. (a) Boundary-loss-dominated and boundary-gain-dominated regimes, $0<\Delta<t_R-t_L$, versus $\Delta$: curves are Eq.~(\ref{simgeneral}). (b) Edge-induced $\mathcal{PT}$-broken regime, $t_R-t_L<\Delta<t_R+t_L$, versus the rescaled coupling $\Delta/(t_R+t_L)$: curves are the approximate local upper envelope Eq.~(\ref{slowerstrong}).}
	\label{imed}
\end{figure}

For $L\ln\lambda\gg1$ and $|q|\ll1$, the $L$-independent term $2X_{1,4}X_{2,3}$ gives a subleading correction (responsible for the even-$L$ residual loss at $\Delta_c$; see Sec.~\ref{sec:threshold}) and can be omitted. Substituting Eq.~(\ref{sbetashiftgen}) into Eq.~(\ref{sobc}) and keeping only the terms proportional to $\lambda^{L+1}$, we obtain
\begin{equation}
	X_{1,3}X_{2,4}\lambda^{L+1}e^{iq(L+1)}
	+
	X_{1,2}X_{3,4}\lambda^{L+1}e^{-iq(L+1)}
	\approx0,
	\label{sdombeq}
\end{equation}
which yields
\begin{equation}
	q
	\approx
	\frac{i}{2(L+1)}
	\ln\!\left[
	\frac{(\lambda-1)^2(t_R+\lambda t_L)(\lambda t_R+t_L)}
	{(\lambda+1)^2(t_R-\lambda t_L)(\lambda t_R-t_L)}
	\right].
	\label{setaformula}
\end{equation}
Expanding the near-zero branch of Eq.~(\ref{sHbeta}) to first order in the small complex shift $q$ gives $E_+\left(\frac{\pi}{2}+q\right)\approx -(t_R+t_L)q$, using the same linear relation between phase shift and energy as in Eqs.~(\ref{sqeven}) and~(\ref{seveneven}). Identifying this branch with the largest imaginary OBC eigenvalue, as supported by Fig.~\ref{spectra}, gives
\begin{equation}
	\max[\mathrm{Im}(E_{\rm OBC})]
	\approx
	-\frac{t_R+t_L}{2(L+1)}
	\ln\!\left[
	\frac{(\lambda-1)^2(t_R+\lambda t_L)(\lambda t_R+t_L)}
	{(\lambda+1)^2(t_R-\lambda t_L)(\lambda t_R-t_L)}\right]
	\label{simgeneral}
\end{equation}
for $0<\Delta<t_R-t_L$, where $\lambda$ is given by Eq.~(\ref{slambdagen}). The largest imaginary OBC eigenvalue is controlled by the small displacement of the zero-energy roots, while the open boundaries dictate that displacement through Eq.~(\ref{sobc}). Eq.~(\ref{simgeneral}) agrees well with the numerical data in Fig.~\ref{imed}(a) over the plotted range. The boundary-scattering interpretation of the logarithmic factor is given in Sec.~\ref{sec:boundscatt}.

The logarithmic factor in Eq.~(\ref{simgeneral}) takes a more elegant form in terms of the variables
\begin{equation}
	a=\ln\lambda,\qquad b=\ln\frac{t_R}{t_L},
	\label{shyperdef}
\end{equation}
where we define $\mathcal R(\lambda)$ by
\begin{equation}
	\mathcal R(\lambda)\equiv
	\frac{(\lambda-1)^2(t_R+\lambda t_L)(\lambda t_R+t_L)}
	{(\lambda+1)^2(t_R-\lambda t_L)(\lambda t_R-t_L)}
	=
	\frac{\cosh a-1}{\cosh a+1}
	\frac{\cosh b+\cosh a}{\cosh b-\cosh a}.
	\label{sRhyper}
\end{equation}
Moreover, Eq.~(\ref{slambdagen}) implies the following relation between $a$ and $b$:
\begin{equation}
	\sinh a
	=
	\frac{(t_R+t_L)V}{2t_Rt_L}
	=
	\frac{V}{\sqrt{t_Rt_L}}
	\cosh\!\left(\frac{b}{2}\right)=\sqrt{\sinh^2b-\frac{\Delta^2}{t_Rt_L}\cosh^2\!\left(\frac{b}{2}\right)}.
	\label{slambdahyper}
\end{equation}
This representation makes the two limiting behaviors transparent: as $\Delta\to0$, $\lambda\to t_R/t_L$ and $\cosh b-\cosh a\to0$ (equivalently $t_R-\lambda t_L\to0$), so $\mathcal R\to\infty$; as $\Delta\to t_R-t_L$, $\lambda\to1$ and $\cosh a-1\to0$ [equivalently $(\lambda-1)^2\to0$], so $\mathcal R\to0$.

Before examining the finite-size gain threshold, we check the small-$\Delta$ limit against the cNHSE scaling in Refs.~\cite{cnhse2,cnhse3}. Expanding for $\Delta\ll t_R-t_L$ gives
\begin{equation}
	\lambda
	=
	\frac{t_R}{t_L}
	-\frac{t_R(t_R+t_L)}{2t_L(t_R-t_L)(t_R^2+t_L^2)}\Delta^2
	+O(\Delta^4).
	\label{slambdaexpand}
\end{equation}
Substituting Eq.~(\ref{slambdaexpand}) into Eq.~(\ref{simgeneral}) yields
\begin{equation}
	\max[\mathrm{Im}(E_{\rm OBC})]
	\approx
	-\frac{t_R+t_L}{L+1}
	\ln\!\left[
	\frac{2(t_R-t_L)(t_R^2+t_L^2)}
	{\Delta(t_R+t_L)^2}
	\right],
	\label{simsmallDelta}
\end{equation}
which recovers the logarithmic dependence on $\Delta$ divided by $L+1$ found in Refs.~\cite{cnhse2,cnhse3}. Eq.~(\ref{simgeneral}) extends this limit across the boundary amplification threshold: $\mathcal R$ decreases through unity at $\Delta_c$, reversing the sign of the spectral rate. Within $1<\lambda<t_R/t_L$, the two logarithmic divergences occur at the coupling endpoints, where the approximation becomes nonuniform in $L$: the small-$q$ condition fails near $\Delta=0$, and $L\ln\lambda\gg1$ fails near $\Delta_{\rm PT}^{\rm PBC}$.

\subsubsection{Convergence of the finite-size OBC gain threshold to \texorpdfstring{$\Delta_c$}{Delta(c)}}\label{sec:exactthreshold}

At $\Delta=\Delta_c$, $\mathcal R=1$ and the leading $L^{-1}$ contribution in Eq.~(\ref{simgeneral}) vanishes, corresponding to perfect reflection. The remaining finite-size loss is exponentially small for even $L$ and asymptotically proportional to $(L+1)^{-2}$ for odd $L$ (Sec.~\ref{sec:threshold}). We denote the leading-order contribution by
\begin{equation}
	M(L,\Delta)
	\equiv
	-\frac{t_R+t_L}{2(L+1)}
	\ln\!\left[
	\frac{(\lambda-1)^2(t_R+\lambda t_L)(\lambda t_R+t_L)}
	{(\lambda+1)^2(t_R-\lambda t_L)(\lambda t_R-t_L)}
	\right],
	\qquad
	M(L,\Delta_c)=0.
	\label{sMdef}
\end{equation}
Set $\Delta=\Delta_c+\delta$ with $|\delta|\ll\Delta_c$. Expanding the leading-order contribution Eq.~(\ref{sMdef}) to first order in $\delta$ gives
\begin{equation}
	M(L,\Delta_c+\delta)
	\simeq
	\frac{2(t_R+t_L)}{V_c(L+1)}\delta,
	\label{sMlinearDelta}
\end{equation}
with $V_c>0$ [Eq.~(\ref{vc})]. The leading $L^{-1}$ term is positive for fixed $\delta>0$. Balancing it against the residual loss estimates the displacement of the finite-size gain threshold.

For even $L$, adding the residual correction in Eq.~(\ref{seveneven}) to Eq.~(\ref{sMlinearDelta}) yields
\begin{equation}
	\max[\mathrm{Im}(E_{\rm OBC})]
	\simeq
	\frac{2(t_R+t_L)}{V_c(L+1)}\left(\delta-V_c\lambda_c^{-L-1}\right).
	\label{sMevendelta}
\end{equation}
To the retained order, this predicts
\begin{equation}
	\max[\mathrm{Im}(E_{\rm OBC})]>0
	\qquad
	{\rm for}\qquad
	\Delta-\Delta_c>V_c\lambda_c^{-L-1}.
	\label{sevencondition}
\end{equation}
The estimated threshold shift on the right-hand side vanishes exponentially with $L$. Therefore, for any fixed $\delta>0$, the positive leading-order term in Eq.~(\ref{sMlinearDelta}) necessarily dominates Eq.~(\ref{seveneven}) once the even system size is sufficiently large.

For odd $L$, using the crossover estimate in Eq.~(\ref{soddthres}), including its phenomenological exponential term, gives
\begin{equation}
	\max[\mathrm{Im}(E_{\rm OBC})]
	\simeq
	\frac{2(t_R+t_L)}{V_c(L+1)}
	\left[
	\delta
	-
	V_c\lambda_c^{-L-1}
	-\frac{(t_R-t_L)^2\pi^2}{16(t_R+t_L)(L+1)}
	\right].
	\label{sModddelta}
\end{equation}
To the retained order, this predicts
\begin{equation}
	\max[\mathrm{Im}(E_{\rm OBC})]>0
	\qquad
	{\rm for}\qquad
	\Delta-\Delta_c>
	V_c\lambda_c^{-L-1}
	+\frac{(t_R-t_L)^2\pi^2}{16(t_R+t_L)(L+1)}.
	\label{soddcondition}
\end{equation}
The odd-$L$ standing-wave phase shift produces an algebraic gain-threshold shift proportional to $(L+1)^{-1}$. Thus, for any fixed $\delta>0$, the leading-order positive term proportional to $\delta/(L+1)$ eventually dominates the odd-$L$ residual correction, whose algebraic part scales as $(L+1)^{-2}$.

Eqs.~\eqref{sevencondition} and \eqref{soddcondition} estimate how the finite-size OBC gain threshold approaches the $L$-independent boundary threshold $\Delta_c$: exponentially for even $L$ and algebraically for odd $L$. For fixed $\Delta_c<\Delta<\Delta_{\rm PT}^{\rm PBC}$, the positive leading $L^{-1}$ rate therefore emerges at sufficiently large $L$.

\subsection{Edge-induced \texorpdfstring{$\mathcal{PT}$}{PT}-broken regime: \texorpdfstring{$\Delta_{\rm PT}^{\rm PBC}<\Delta<\Delta_{\rm PT}^{\rm OBC}$}{Delta(PT, PBC) < Delta < Delta(PT, OBC)}}\label{sec:strongcoupl}

In the edge-induced $\mathcal{PT}$-broken regime, $t_R-t_L<\Delta<t_R+t_L$, $V=0$ and the PBC spectrum is purely real. For a given finite chain, complex OBC branches occur only where allowed by boundary quantization, with imaginary parts that vary nonmonotonically with coupling (Fig.~\ref{spectra}). We derive the near-zero branch for even $L$ and the resulting local upper-envelope estimate. At $E_{\rm OBC}=0$ all four roots of $\det[H(\beta)-0]=0$ lie on the unit circle, with reciprocal pairing $\beta_1\beta_4=\beta_2\beta_3=1$ and phases that can be parameterized as $\pi/2\pm p$ and $-\pi/2\pm p$. A small purely imaginary displacement of $E_{\rm OBC}$ moves these roots off the unit circle while preserving their reciprocal pairing. We write
\begin{equation}
	\beta_1\approx e^{-\kappa+i(\pi/2+p)},\quad
	\beta_2\approx e^{-\kappa+i(\pi/2-p)},\quad
	\beta_3\approx e^{\kappa-i(\pi/2-p)},\quad
	\beta_4\approx e^{\kappa-i(\pi/2+p)} ,
\end{equation}
where $0<p<\pi/2$ is the real phase displacement around $\pi/2$ and $\kappa>0$ measures the small departure from the unit circle. The characteristic equation can be written as
\begin{equation}
t_Rt_L\beta^4-(t_R+t_L)E_{\rm OBC}\beta^3+
(E_{\rm OBC}^2+t_R^2+t_L^2-\Delta^2)\beta^2
-(t_R+t_L)E_{\rm OBC}\beta+t_Rt_L=0 .
\end{equation}
Using the Vieta relations for the above roots, we obtain
\begin{equation}
	E_{\rm OBC}=-\frac{4it_Rt_L}{t_R+t_L}\cos p\,\sinh\kappa,
	\label{sstrongE}
\end{equation}
together with
\begin{equation}
	\Delta^2=\frac{(t_R^2+t_L^2-2t_Rt_L\cos2p)(t_R^2+t_L^2+2t_Rt_L\cosh2\kappa)}{(t_R+t_L)^2}.
	\label{sstrongDeltaExact}
\end{equation}
Where this branch exists at fixed coupling away from the endpoints, $\kappa\ll1$ for large $L$. Expanding Eq.~(\ref{sstrongDeltaExact}) to leading order in $\kappa$, we find
\begin{equation}
	\sin^2 p\approx \frac{\Delta^2-(t_R-t_L)^2}{4t_Rt_L},
	\qquad
	\cos^2 p\approx \frac{(t_R+t_L)^2-\Delta^2}{4t_Rt_L}.
	\label{sstrongp}
\end{equation}
These expressions are well defined precisely in the interval $t_R-t_L<\Delta<t_R+t_L$. Eq.~(\ref{sstrongE}) gives one member of a complex-conjugate pair; the partner, equivalently the opposite sign of $\kappa$, has the opposite imaginary part. Since we are interested in $\max[\mathrm{Im}(E_{\rm OBC})]$, we take the positive-imaginary branch in the following. Thus,
\begin{equation}
	\mathrm{Im}(E_{\rm OBC})
	\approx
	\frac{2\sqrt{t_Rt_L\left[(t_R+t_L)^2-\Delta^2\right]}}{t_R+t_L}\,\kappa.
	\label{sstrongime0}
\end{equation}
Therefore the remaining task is to determine $\kappa$ from the OBC boundary equation.

For even $L$, expanding each $X_{i,j}$ in Eq.~(\ref{sobc}) to leading order in $\kappa$ while keeping the products $(L+1)\kappa$ nonperturbative, substituting the parametrization above, and using Eq.~(\ref{sstrongp}), gives the boundary constraint
\begin{equation}
	4t_Rt_L\Delta^2\cos[2(L+1)p]
	-(t_R-t_L)^2\left[(t_R+t_L)^2-\Delta^2\right]
	+(t_R+t_L)^2\left[\Delta^2-(t_R-t_L)^2\right]\cosh[2(L+1)\kappa]
	\approx0.
	\label{sstrongsobc2}
\end{equation}
Solving this equation for $\kappa$ gives a real value only when the argument of $\operatorname{arcosh}$ is not smaller than one:
\begin{equation}
	\kappa
	\approx
	\frac{1}{2(L+1)}
	\operatorname{arcosh}\!\left[
	\frac{(t_R-t_L)^2\big[(t_R+t_L)^2-\Delta^2\big]-4t_Rt_L\Delta^2\cos[2(L+1)p]}
	{(t_R+t_L)^2\big[\Delta^2-(t_R-t_L)^2\big]}
	\right].
	\label{sstrongkappa}
\end{equation}
Within this even-$L$ approximation, an argument greater than unity gives a purely imaginary pair with $\kappa>0$; at unity, the pair reaches $E=0$. This condition concerns the near-zero branch and does not determine whether other complex eigenvalues exist.
According to Eq.~(\ref{sstrongime0}), $\mathrm{Im}(E_{\rm OBC})$ is proportional to $\kappa$. Since Eq.~(\ref{sstrongp}) fixes $p$ as a function of $\Delta$, the factor $\cos[2(L+1)p(\Delta)]$ in Eq.~(\ref{sstrongkappa}) oscillates rapidly as $\Delta$ is swept. Treating the remaining coefficients as slowly varying over one oscillation cycle, replacing $\cos[2(L+1)p]$ by $-1$ gives the local upper-envelope estimate
\begin{equation}
	\left.\max[\mathrm{Im}(E_{\rm OBC})]\right|_{\rm envelope}
	\approx
	\frac{\sqrt{t_Rt_L\big[(t_R+t_L)^2-\Delta^2\big]}}{(t_R+t_L)(L+1)}
	\operatorname{arcosh}\!\left[
	\frac{(t_R-t_L)^2\big[(t_R+t_L)^2-\Delta^2\big]+4t_Rt_L\Delta^2}
	{(t_R+t_L)^2\big[\Delta^2-(t_R-t_L)^2\big]}
	\right].
	\label{slowerstrong}
\end{equation}
Eq.~(\ref{slowerstrong}) predicts an $L^{-1}$ envelope at fixed coupling away from the endpoints. Its coupling dependence agrees with the numerical envelope for odd $L=81$ in Fig.~\ref{imed}(b), supporting its use beyond the even-$L$ near-zero branch derived here. The apparent divergence near $\Delta_{\rm PT}^{\rm PBC}$ is rounded at finite $L$, while both PBC and OBC spectra are real for $\Delta\geq\Delta_{\rm PT}^{\rm OBC}$.

\section{Boundary round-trip scattering theory of NHEA}\label{sec:boundscatt}

While the finite-size spectral result in Eq.~(\ref{simgeneral}) establishes that an open boundary can generate net amplification, it does not reveal the underlying physical mechanism. In the NHEA regime ($0<\Delta<\Delta_{\rm PT}^{\rm PBC}=t_R-t_L$), the on-site loss $V$ pins the bulk reference to $\max[\operatorname{Im}(E_{\rm PBC})]=0$, so bulk Bloch waves provide no exponential growth. The open-boundary amplification must therefore originate entirely from boundary reflections, which impart an $L$-independent norm gain per reflection event. To understand how an open boundary transforms from an absorber into an amplifier, we formulate a boundary-scattering theory within the non-Bloch framework, analyzing first the local reflection at an isolated semi-infinite boundary and then imposing round-trip consistency across the finite chain.

Because the leading open-boundary eigenvalue approaches zero as $E_{\rm OBC}^{\max}\sim\mathcal{O}(L^{-1})$ (Sec.~\ref{sec:approxime}\,A), local boundary matching at an isolated wall is asymptotically exact at the reference energy $E=0$, with finite-size corrections entering only through spatial transit across the bulk. At $E=0$, the non-Bloch characteristic equation $\det[H(\beta)]=0$ yields four roots:
\begin{equation}
	\beta_1^{(0)}=-i/\lambda,\qquad
	\beta_2^{(0)}=+i,\qquad
	\beta_3^{(0)}=-i,\qquad
	\beta_4^{(0)}=+i\lambda,
	\label{appReferenceRoots}
\end{equation}
with $1<\lambda<t_R/t_L$ given by Eq.~(\ref{slambdagen}). These roots partition cleanly into two physical sectors: the unit-modulus pair $\beta_{2,3}^{(0)}=\pm i$ defines counterpropagating bulk channels with group velocities $\mp v=\mp(t_R+t_L)$, whereas the off-unit-circle roots $\beta_1^{(0)}$ and $\beta_4^{(0)}$ describe skin modes evanescent from the left and right boundaries with decay length $1/\ln\lambda$.

At reference energy $E=0$, each root $\beta_j^{(0)}$ is uniquely paired with the internal sublattice spinor in the kernel $H(\beta_j^{(0)})\phi(\beta_j^{(0)})=0$. Evaluating the band dispersion establishes the inter-branch assignments:
\begin{equation}
	E_+(\beta_2^{(0)})=E_+(\beta_3^{(0)})
	=E_-(\beta_1^{(0)})=E_-(\beta_4^{(0)})=0.
	\label{appReferencePairing}
\end{equation}
The alternative assignments are excluded because their eigenvalues are strictly nonzero for $V>0$, e.g., $E_-(\beta_{2,3}^{(0)})=-2iV$. Each root is thus uniquely paired with a zero-energy spinor. Suppressing reference superscripts below, we formulate boundary reflection using these four zero-energy spinors.

To analyze boundary reflection, we adopt the two normalized propagation spinors as the scattering channel basis:
\begin{equation}
	\phi_R\equiv\phi(\beta_3)=
	\begin{pmatrix}
		\sqrt{\dfrac{t_R-t_L+V}{2(t_R-t_L)}}\\
		-i\sqrt{\dfrac{t_R-t_L-V}{2(t_R-t_L)}}
	\end{pmatrix},\qquad
	\phi_L\equiv\phi(\beta_2)=
	\begin{pmatrix}
		\sqrt{\dfrac{t_R-t_L-V}{2(t_R-t_L)}}\\
		i\sqrt{\dfrac{t_R-t_L+V}{2(t_R-t_L)}}
	\end{pmatrix}.
	\label{appReferencePropSpinors}
\end{equation}
Crucially, these zero-energy propagation spinors are strictly orthogonal, $\phi_R^\dagger\phi_L=0$. Consequently, direct reflection between propagating channels cannot satisfy the boundary condition $\psi_0=0$ (or $\psi_{L+1}=0$) on its own; boundary matching must be mediated by the boundary-localized skin modes. The transformation matrix $U_\Delta\equiv(\phi_R,\phi_L)$ is therefore unitary, mapping the propagating spinors to canonical basis vectors $U_\Delta^\dagger\phi_R=(1,0)^T$ and $U_\Delta^\dagger\phi_L=(0,1)^T$. The components of a transformed skin spinor, $\tilde\phi(\beta_j)\equiv U_\Delta^\dagger\phi(\beta_j)$, directly represent its projection amplitudes onto the propagating channels.

Normalizing the zero-energy skin solutions of $H(\beta_{1,4}^{(0)})\phi=0$ yields:
\begin{equation}
	\begin{aligned}
		\phi(\beta_4)&=\frac{1}{\sqrt{\Delta^2+(V-t_L\lambda+t_R/\lambda)^2}}
		\begin{pmatrix}\Delta\\i(V-t_L\lambda+t_R/\lambda)\end{pmatrix},\\
		\phi(\beta_1)&=\frac{1}{\sqrt{\Delta^2+(V-t_R\lambda+t_L/\lambda)^2}}
		\begin{pmatrix}\Delta\\i(V-t_R\lambda+t_L/\lambda)\end{pmatrix}.
	\end{aligned}
	\label{appReferenceSkinSpinors}
\end{equation}
Reflection converts an incident left-moving mode ($\phi_L$) into an outgoing right-moving mode ($\phi_R$) at the left wall ($x=0$) via $\phi(\beta_1)$, and vice versa at the right wall ($x=L+1$) via $\phi(\beta_4)$. In the channel basis, choosing suitable overall phases parameterizes these skin spinors by a single geometric overlap angle $\theta\in(0,\pi/2)$:
\begin{equation}
	\tilde\phi(\beta_4)=\begin{pmatrix}\cos\theta\\\sin\theta\end{pmatrix},
	\qquad
	\tilde\phi(\beta_1)=\begin{pmatrix}-\sin\theta\\\cos\theta\end{pmatrix},
	\qquad \theta\in(0,\pi/2).
	\label{appSkin}
\end{equation}
Both skin spinors thus exhibit identical overlap geometry: each has projection magnitude $\cos\theta=|\phi_L^\dagger\phi_1|=|\phi_R^\dagger\phi_4|$ onto its respective incident channel and $\sin\theta$ onto its outgoing channel.

Direct evaluation of the projection ratio $\cot^2\theta=\cos^2\theta/\sin^2\theta$ yields:
\begin{equation}
	\cot^2\theta=\mathcal R(\lambda)\equiv
	\frac{(\lambda-1)^2(t_R+\lambda t_L)(\lambda t_R+t_L)}
	{(\lambda+1)^2(t_R-\lambda t_L)(\lambda t_R-t_L)}.
	\label{appCotR}
\end{equation}
Eq.~(\ref{appCotR}) establishes the fundamental bridge between boundary geometry and spectral flow: across the entire NHEA regime where $1<\lambda<t_R/t_L$, it uniquely identifies the non-Bloch spectral factor $\mathcal{R}(\lambda)$ from Eq.~(\ref{simgeneral}) as the ratio of the squared magnitudes of the incident and outgoing skin-mode projections, fixing the overlap angle $\theta\in(0,\pi/2)$.

To determine how each boundary reflects propagating waves, we expand the state in the channel basis as $\tilde\Psi_x=\sum_{j}\tilde c_j\beta_j^x\tilde\phi(\beta_j)$ [generalizing Eq.~(\ref{sgbz})]. In the semi-infinite limit at an isolated wall, physical boundary states retain only the propagating channels ($\beta_{2,3}$) and the skin mode that decays exponentially into the bulk: $\beta_1$ for the left wall ($|\beta_1|=1/\lambda<1$) and $\beta_4$ for the right wall ($|\beta_4|=\lambda>1$). Enforcing the open boundary conditions $\tilde\Psi_0=0$ and $\tilde\Psi_{L+1}=0$ [Eq.~(\ref{obc})] then casts the wall matching into algebraic vector form:
\begin{equation}
	\tilde c_1\begin{pmatrix}-\sin\theta\\ \cos\theta\end{pmatrix}
	+\tilde c_2\begin{pmatrix}0\\ 1\end{pmatrix}
	+\tilde c_3\begin{pmatrix}1\\ 0\end{pmatrix}=0
	\qquad(\text{left wall},\ x=0),
	\label{appBCleft}
\end{equation}
\begin{equation}
	\tilde c_2\beta_2^{L+1}\begin{pmatrix}0\\ 1\end{pmatrix}
	+\tilde c_3\beta_3^{L+1}\begin{pmatrix}1\\ 0\end{pmatrix}
	+\tilde c_4\beta_4^{L+1}\begin{pmatrix}\cos\theta\\ \sin\theta\end{pmatrix}=0
	\qquad(\text{right wall},\ x=L+1).
	\label{appBCright}
\end{equation}
Eliminating the mediating skin amplitude at each wall directly yields the reflection amplitude ratios:
\begin{equation}
	r_L\equiv\frac{\tilde c_3}{\tilde c_2}=-\tan\theta,
	\qquad
	r_R\equiv\frac{\tilde c_2\beta_2^{L+1}}{\tilde c_3\beta_3^{L+1}}=\tan\theta.
	\label{appRtangen}
\end{equation}
The underlying microscopic mechanism is transparent: canceling a unit-amplitude incident wave along the incoming channel requires exciting the boundary skin mode with amplitude $1/\cos\theta$; the outgoing projection of this induced skin component then inevitably emits a reflected wave of amplitude $(1/\cos\theta)\sin\theta=\tan\theta$. Consequently, whether an isolated boundary attenuates ($|r_{L,R}|<1$) or amplifies ($|r_{L,R}|>1$) the incident wave is governed entirely by whether the geometric projection angle $\theta$ is smaller or larger than $\pi/4$.

For a wave packet initialized at the chain center [$x_0=(L+1)/2$], the counterpropagating components reach opposite walls simultaneously. Each packet acquires a norm multiplier $r_L^2=r_R^2=\tan^2\theta$, so that their combined norm scales by the exact same factor, establishing
\begin{equation}
	\mathcal R(\lambda)=\frac{1}{r_L^2}=\frac{1}{r_R^2}.
	\label{appRfromrr}
\end{equation}
The non-Bloch spectral factor $-\ln\mathcal R(\lambda)=\ln\tan^2\theta$ is thus precisely the logarithmic norm gain imparted by each paired reflection encounter.

In a finite chain, these local reflections must be reconciled with bulk propagation to determine the open-boundary spectrum. When the skin modes decaying from opposite walls have negligible mutual overlap ($L\ln\lambda\gg1$), the local reflection laws remain asymptotically exact at $E=0$. Although the finite-size momentum shift $q$ in $\beta_2/\beta_3\simeq -e^{2iq}$ is small ($|q|\sim\mathcal{O}(L^{-1})$), the spatial factor $(L+1)q$ accumulated across the chain is of order unity and dictates the round-trip boundary quantization. Matching the reflection conditions at both walls [Eq.~(\ref{appRtangen})] requires $r_Lr_R=(\beta_2/\beta_3)^{L+1}\simeq -\tan^2\theta$, yielding the round-trip quantization condition:
\begin{equation}
	e^{2iq(L+1)}\simeq(-1)^L\tan^2\theta.
	\label{appFP}
\end{equation}
For even $L$, where $q$ is purely imaginary near zero energy (Sec.~\ref{sec:approxime}\,A), substituting $\tan^2\theta=\mathcal R^{-1}(\lambda)$ [Eq.~(\ref{appRfromrr})] yields the quantized shift:
\begin{equation}
	q\simeq\frac{i}{2(L+1)}\ln\mathcal R(\lambda).
	\label{appq}
\end{equation}
Linearizing the band dispersion near $k=\pi/2$, $E_+(\pi/2+q)\simeq-vq$ (Sec.~\ref{sec:approxime}\,A), directly recovers the dominant open-boundary growth rate:
\begin{equation}
	\max[\mathrm{Im}(E_{\rm OBC})]\simeq
	-\frac{v}{2(L+1)}\ln\mathcal R(\lambda).
	\label{appImE}
\end{equation}
This reproduces the leading near-zero-branch expression in Eq.~(\ref{simgeneral}), whose agreement with numerical diagonalization is confirmed in Fig.~\ref{imed}(a), directly tracing the open-boundary growth rate to the norm gain imparted by boundary reflections.

Removing the bulk decay background reveals the discrete logarithmic step imparted by each paired reflection:
\begin{equation}
	\delta h\simeq-\ln\mathcal R(\lambda)=\ln\tan^2\theta.
	\label{appStepReflection}
\end{equation}
Because consecutive boundary encounters are separated by the transit time $T\equiv(L+1)/v$, the discrete step connects directly to the continuous spectral growth rate in Eq.~(\ref{appImE}) via
\begin{equation}
	\delta h\simeq 2T\max[\operatorname{Im}(E_{\rm OBC})].
	\label{appStepSpectral}
\end{equation}
This identity reconciles the discrete dynamics with continuous non-Bloch spectral flow: while local reflection fixes an $L$-independent boundary step $\delta h\sim\mathcal{O}(1)$, the transit time grows as $T\propto L$, forcing the continuous spectral rate $\max[\operatorname{Im}(E_{\rm OBC})]\sim\mathcal{O}(L^{-1})$ to vanish in the thermodynamic limit.

This framework provides an intuitive geometric origin for the boundary amplification threshold $\Delta_c$ (Sec.~\ref{sec:threshold}). Norm-preserving reflection ($r_L^2=r_R^2=1$, or $\delta h\simeq 0$) requires the mediating skin spinor to project equally onto the incoming and outgoing channels ($\cos\theta=\sin\theta=1/\sqrt{2}$), fixing $\theta=\pi/4$. Setting $\cot^2(\pi/4)=1$ in Eq.~(\ref{appCotR}) directly recovers $\Delta_c=(t_R-t_L)\sqrt{t_R^2+t_L^2}/(t_R+t_L)$, showing that boundary loss ($\delta h<0$) and NHEA gain ($\delta h>0$) correspond to $\theta<\pi/4$ and $\theta>\pi/4$, respectively.

\begin{figure}[!t]
	\centering
	\includegraphics[width=0.4\textwidth]{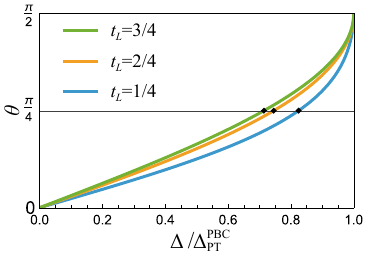}
	\caption{Geometric overlap angle $\theta$ as a function of the normalized coupling $\Delta/\Delta_{\rm PT}^{\rm PBC}$ for $t_L/t_R=1/4, 2/4, 3/4$ with $t_R=1$. The horizontal line marks $\theta=\pi/4$, representing norm-preserving reflection ($r_L^2=r_R^2=1$), with dots indicating the boundary amplification threshold $\Delta_c$. Across the physical interval, $\theta$ monotonically increases from $0$ (absorbing boundary of conventional NHSE) to $\pi/2$ (bulk exceptional point mode coalescence).}\label{fig:theta}
\end{figure}

As the coupling $\Delta$ increases across $(0, \Delta_{\rm PT}^{\rm PBC})$, the overlap angle $\theta$ monotonically enlarges from $0$ to $\pi/2$ [Fig.~\ref{fig:theta}], continuously connecting two distinct physical regimes. In the decoupled limit $\Delta\to0^+$, the skin mode aligns with the incident channel, yielding $\theta\to0$ ($\delta h\to-\infty$) and recovering the absorbing boundary behavior of the conventional non-Hermitian skin effect~\cite{cnhse2,cnhse3}. Conversely, at the upper boundary $\Delta\to(\Delta_{\rm PT}^{\rm PBC})^-$, $\theta$ approaches $\pi/2$, driving a formal reflection divergence ($\delta h\to+\infty$) whose microscopic origin lies in the bulk exceptional points.

In the scattering channel basis, the Hamiltonian at the two channel roots $\beta=\mp i$, $H_r(\beta)\equiv U_\Delta^\dagger H(\beta)U_\Delta$, takes the form
\begin{equation}
	H_r(-i)=
	\begin{pmatrix}
		0&2i\Delta\\
		0&-2iV
	\end{pmatrix},
	\qquad
	H_r(+i)=
	\begin{pmatrix}
		-2iV&0\\
		-2i\Delta&0
	\end{pmatrix}.
	\label{appHrotpoles}
\end{equation}
At $\Delta=\Delta_{\rm PT}^{\rm PBC}\equiv t_R-t_L$, $V=0$ collapses each matrix into a defective $2\times2$ Jordan block. Approaching this exceptional point ($\Delta\to(\Delta_{\rm PT}^{\rm PBC})^-$), the skin roots merge with the propagating channels ($\beta_4\to\beta_2=+i$ and $\beta_1\to\beta_3=-i$), and the skin spinors rotate into collinearity with the outgoing channels ($\cos\theta\to 0$). Canceling the incoming wave then requires an excited skin amplitude $\propto 1/\cos\theta\to\infty$, emitting a diverging reflected wave $\tan\theta\to\infty$. The diverging step $\delta h\simeq\ln\tan^2\theta$ is thus the boundary-scattering signature of bulk eigenvector coalescence.

In a finite chain, this formal divergence is regularized once the skin decay length $\xi=1/\ln\lambda$ becomes comparable to the system size ($L\ln\lambda\lesssim 1$). The skin tails from opposite boundaries then overlap across the bulk, breaking the semi-infinite decoupling and rounding off the singular reflection into a finite maximum gain governed by Eq.~(\ref{sobc}). The semi-infinite scattering theory thus captures the reflection mechanism throughout $L\ln\lambda\gg 1$, while boundary overlap regularizes the critical endpoint.

\section{Dynamical behavior of the NHEA walker}\label{sec:walkerdynamics}

\subsection{Decay behavior of the walker in the bulk}\label{sec:bulkdecay}

    The $t^{-1/2}$ bulk decay provides the reference envelope against which boundary attenuation and amplification are measured. For $0\leq\Delta<\Delta_{\rm PT}^{\rm PBC}=t_R-t_L$, we derive this envelope from the PBC dispersion using an asymptotic expansion about its zero-growth momenta. The resulting power law describes the intermediate-time bulk evolution before boundary encounters become important. Before the packets reach a boundary, the same bulk propagation governs OBC dynamics~\cite{pnhse,bcinhd}. At diagonalizable momenta, the biorthogonal spectral decomposition $H(k)=\sum_{l=\pm}E_l(k)|\psi_{Rkl}\rangle\langle \psi_{Lkl}|$ with $\langle\psi_{Lkl}|\psi_{Rkl'}\rangle=\delta_{ll'}$, combined with $|x_0,B\rangle=L^{-1/2}\sum_k e^{-ikx_0}|k,B\rangle$ and $\langle k'|k\rangle=\delta_{kk'}$, yields
    \begin{equation}
		\begin{split}
		\mathcal{N}(t)=&\langle \psi(t)|\psi(t) \rangle
             \\=&\langle x_0,B|e^{iH^\dagger t}e^{-iHt}|x_0,B\rangle \\ =&\frac{1}{L}\sum_{k,ll'} e^{iE_l^*(k)t-iE_{l'}(k)t}(|\psi_{Lkl}\rangle\langle \psi_{Rkl}|\psi_{Rkl'}\rangle\langle \psi_{Lkl'}|)_{BB},
		\end{split}
            \label{spsit}
    \end{equation}
    where $(\cdots)_{BB}\equiv\langle B|\cdots|B\rangle$. At exceptional momenta, the complete eigenstate sum in Eq.~(\ref{spsit}) is understood by continuity. For fixed $V>0$, the dominant momenta $k=\pm\pi/2$ are diagonalizable, so the saddle-point expansion below is unaffected. The cross factor $\langle\psi_{Rkl}|\psi_{Rkl'}\rangle$ does not collapse to $\delta_{ll'}$ because the right eigenstates of a non-Hermitian Bloch Hamiltonian are not mutually orthogonal.

    In the window $0\le\Delta<t_R-t_L$, ${\rm max} \{ {\rm Im}[E_{\pm}(k)] \}={\rm Im}[E_+(\pm \pi/2)]=0$ while $\mathrm{Im}[E_-(k)]<0$ generically. In the continuum regime, the momentum integral is dominated by the two zero-growth momenta $k=\pm\pi/2$ and the $l=l'=+$ sector. With $\Delta_{\rm PT}^{\rm PBC}\equiv t_R-t_L$ and $v\equiv t_R+t_L$, expansion of $E_+(k)$ around $k=\pm\pi/2+\delta k$ gives
    \begin{equation}
    	E_+(\pm\pi/2+\delta k)\approx \mp v\,\delta k-i\frac{(\Delta_{\rm PT}^{\rm PBC})^2}{2V}(\delta k)^2,
    	\label{Eexpand}
    \end{equation}
	so $\mathrm{Im}[E_+^*-E_+]=(\Delta_{\rm PT}^{\rm PBC})^2(\delta k)^2/V$. For the initial state $|B\rangle$, the two zero-growth momenta carry different projection weights. Their average, which enters the Gaussian integral, is
	\begin{equation}
		\frac12\sum_{\sigma=\pm1}\left.(|\psi_{Lk+}\rangle\langle\psi_{Rk+}|\psi_{Rk+}\rangle\langle\psi_{Lk+}|)_{BB}\right|_{k=\sigma\pi/2}
		=
		\frac{(\Delta_{\rm PT}^{\rm PBC})^2}{2((\Delta_{\rm PT}^{\rm PBC})^2-\Delta^2)},
	\end{equation}
	The identical quadratic curvature at the two momenta then gives
    \begin{equation}
    \begin{split}
    \mathcal N(t)
    \approx&
    \frac{1}{\pi}\int d(\delta k)\,
    \frac{(\Delta_{\rm PT}^{\rm PBC})^2}{2\left((\Delta_{\rm PT}^{\rm PBC})^2-\Delta^2\right)}
    \exp\!\left[-\frac{(\Delta_{\rm PT}^{\rm PBC})^2}{V}\,(\delta k)^2 t\right] \\
    =&
    \frac{\Delta_{\rm PT}^{\rm PBC}}{2\sqrt{\pi}\,\left((\Delta_{\rm PT}^{\rm PBC})^2-\Delta^2\right)^{3/4}}\,t^{-1/2}
    \sim t^{-1/2},
    \end{split}
    \label{spsit1}
    \end{equation}
    The quadratic maximum of the imaginary dispersion fixes the bulk exponent $1/2$, while its curvature and the initial-state projections determine the $\Delta$-dependent prefactor. This continuum approximation applies at fixed $V>0$ once the zero-growth neighborhoods dominate and before finite momentum spacing is resolved.

\subsection{Spatial decay exponent of walker dissipation at the perfect reflection threshold \texorpdfstring{$\Delta=\Delta_c$}{Delta = Delta(c)}}
\begin{figure}[t]
	\centering
	\includegraphics[width=0.7\textwidth]{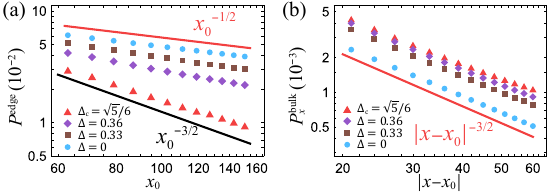}\\
	\caption{(a) Signed accumulated left-edge dissipation, $P^\text{edge}\equiv\sum_{x=1}^{l}P_x$, versus $x_0$ for couplings from $\Delta=0$ to $\Delta_c$, where $l$ specifies the left boundary region. We use $l=16$, where $|P_l|$ is already negligible; the extracted scaling is insensitive to modest changes in this cutoff. The fitted slope gives an effective edge exponent $\alpha_e$, which crosses over from the absorbing-edge value $1/2$ toward the critical reflective-edge value $3/2$ as $\Delta$ increases. (b) Distribution of $P_x$ in the bulk for the same set of $\Delta$ at $x_0=151$. In all panels, $t_R=1$, $t_L=1/2$, and $L=2x_0-1$ so the walker is initialized at the midpoint.}
	\label{figedgeexp}
\end{figure}

At $\Delta=\Delta_c$, reflection preserves the incident wave-packet norm to leading order [Sec.~\ref{sec:boundscatt}], removing the leading absorption contribution that produces the conventional edge burst. We use this boundary balance to estimate the spatial exponent of the signed accumulated edge dissipation and compare it with Fig.~\ref{figedgeexp}. The bulk decay $\mathcal N(t)\sim t^{-1/2}$ derived in the preceding subsection [Eq.~(\ref{spsit1})], together with the zero-growth group velocity $|v_\leftarrow|=t_R+t_L$, relates the temporal decay to the propagation distance. For the accumulated dissipation defined in the End Matter, we take $t_f\to\infty$ at each finite $L$ before varying $x_0$, with $L=2x_0-1$ for midpoint initialization.

Let $P^\text{edge}$ denote the time-integrated dissipation accumulated in the left boundary region. As a reference, at $\Delta=0$ the two chains are decoupled and the walker drifts unidirectionally to the left edge with velocity $-|v_\leftarrow|$; once it arrives at $t\simeq x_0/|v_\leftarrow|$ it is trapped there, and the on-site loss subsequently dissipates the trapped population to $\mathcal N(\infty)=0$. The integrated edge dissipation therefore equals the remaining norm to leading order at the arrival time,
\begin{eqnarray}
	P^\text{edge}|_{\Delta=0}
	&\approx&
	\mathcal{N}(t=x_0/|v_\leftarrow|)-\mathcal{N}(\infty) \notag\\
	&\sim& t^{-1/2}\propto x_0^{-1/2},
	\label{pedge1}
\end{eqnarray}
the conventional edge-burst scaling with exponent $\alpha_e=1/2$.

At $\Delta=\Delta_c$, the walker returns to the bulk after reflection and repeatedly revisits the boundary. To connect the norm envelope to link dissipation, consider an isolated bulk packet $\psi_x\simeq f_x(t)e^{ik_0x}\phi$, where $k_0=\pm\pi/2$, $\phi$ is the corresponding normalized spinor, and $f_x$ varies slowly. The zero-growth condition gives $2V+2(t_R-t_L)\sin k_0\,\phi^\dagger\sigma_z\phi=0$, with $\sigma_z=\operatorname{diag}(1,-1)$ in the $A,B$ basis. Substituting this packet into Eqs.~(11)--(13) in the End Matter, the instantaneous interior-link dissipation in the fixed-spinor approximation is
\begin{equation}
 \frac{dP_x}{dt}\simeq V\bigl(|f_x|^2+|f_{x+1}|^2-2\operatorname{Re}[f_x^*f_{x+1}]\bigr)=V|f_{x+1}-f_x|^2.
 \label{eq:bulk-link-envelope}
\end{equation}
Thus the leading density terms cancel locally. For a packet of width $w\propto\sqrt t$ and norm $\mathcal N_{\rm packet}\propto t^{-1/2}$, integrating over its passage time $w/v$ gives $P_x^{\rm bulk,pass}\sim\mathcal N_{\rm packet}/(vw^2)\propto t^{-3/2}$. The single-passage contribution therefore has the same temporal power as $-d\mathcal N/dt$:
\begin{equation}
	P_x^{\rm bulk,pass}
	\sim
	-\left.
	\frac{d\mathcal{N}(t)}{dt}
	\right|_{t=\frac{|x-x_0|}{|v_\leftarrow|}}
	\sim
	\left(\frac{|x-x_0|}{|v_\leftarrow|}\right)^{-3/2}
	\propto |x-x_0|^{-3/2}.
	\label{pxb1}
\end{equation}
At the threshold, norm-preserving reflection cancels the leading reflected-versus-incident norm loss. To estimate the remaining signed contribution from the $n$th encounter with the chosen boundary region, we assume a derivative response, $P_{\rm edge}^n\propto-d\mathcal N/dt|_{t_n}$, with coefficients that do not scale with the arrival time $t_n$. For midpoint initialization, the two packets reach the left edge alternately at $t_n\simeq(2n-1)x_0/|v_\leftarrow|$ ($n=1,2,\dots$), with successive arrivals separated by approximately $2x_0/|v_\leftarrow|$. Hence
\begin{equation}
	P_{\rm edge}^n\sim [(2n-1)x_0]^{-3/2}.
	\label{pedge_n}
\end{equation}
Let $n_*$ denote the last well-separated boundary encounter, before the packet width $w\propto\sqrt{t_n}$ becomes comparable to $L$. Summing these encounters gives
\begin{equation}
 P^{\rm edge}_{\rm packet}\big|_{\Delta=\Delta_c}
 =\sum_{n=1}^{n_*}P_{\rm edge}^n
 \sim x_0^{-3/2}\sum_{n=1}^{n_*}(2n-1)^{-3/2}
 \sim x_0^{-3/2}.
 \label{pedge2}
\end{equation}
The first few encounters dominate the convergent sum, giving an $x_0^{-3/2}$ edge contribution. The encounter estimate thus predicts matching edge and bulk spatial decay exponents, denoted by $\alpha_e$ and $\alpha_b$, respectively:
\begin{equation}
	\alpha_e=\alpha_b=\frac{3}{2}.
	\label{EA1}
\end{equation}
For $0<\Delta<\Delta_c$, the fitted $\alpha_e$ is an effective crossover exponent over the sampled $x_0$ range. The boundary-scattering picture of Sec.~\ref{sec:boundscatt} shows that increasing $\Delta$ enhances the reflected wave-packet amplitude: weak reflection at small $\Delta$ leaves the result close to the absorbing-edge value $\alpha_e\simeq1/2$, whereas nearly norm-preserving reflection near $\Delta_c$ suppresses this leading absorption and can expose the derivative-response contribution in a finite fitting window. The fitted $\alpha_e$ increases toward $3/2$ in Fig.~\ref{figedgeexp}(a). The local cancellation in Eq.~(\ref{eq:bulk-link-envelope}) and the quadratic dispersion maximum fix the single-passage bulk exponent at $3/2$ throughout this interval. Figure~\ref{figedgeexp}(b) compares the accumulated bulk profiles with this scaling.

  \subsection{Coherent boundary interference of walker wave packets at \texorpdfstring{$\Delta=\Delta_c$}{Delta = Delta(c)}}

\begin{figure}[t]
		\centering
		\includegraphics[width=0.9\textwidth]{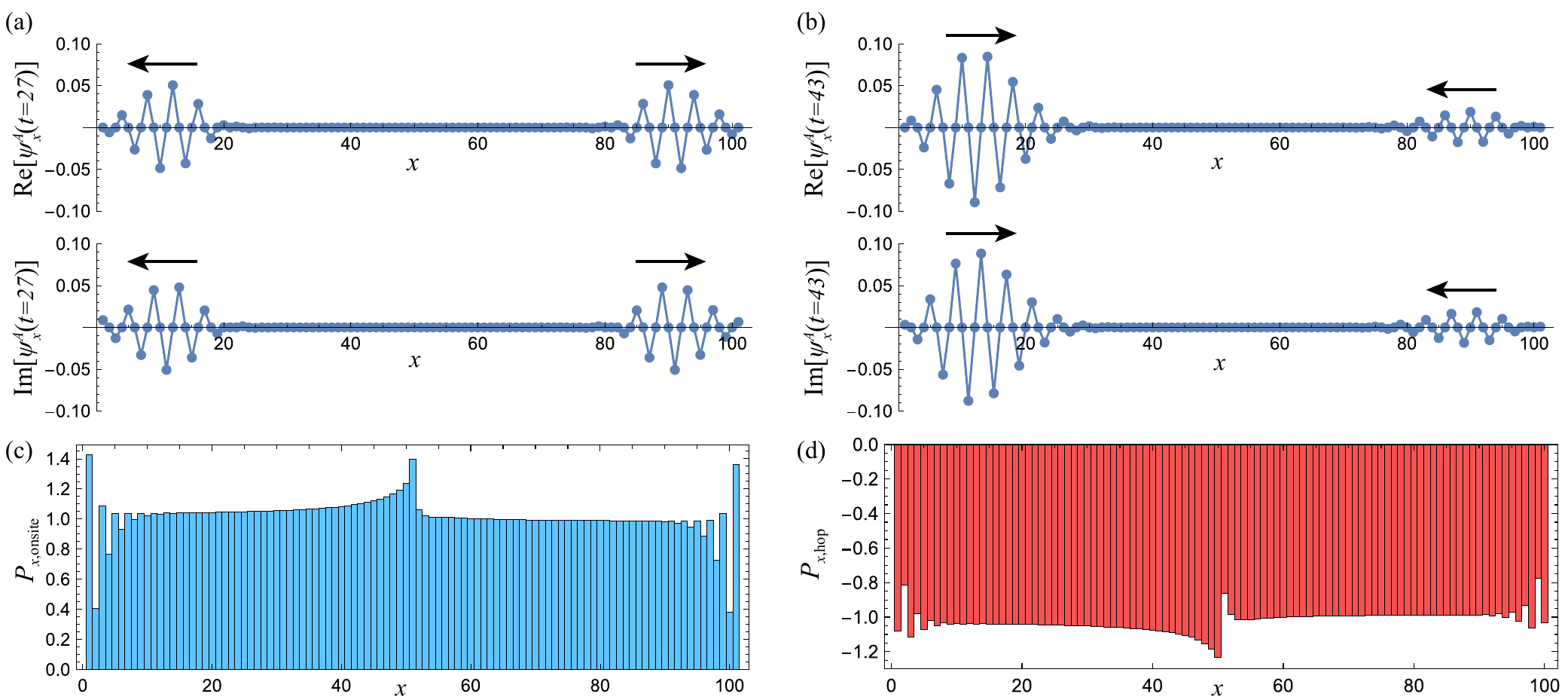}\\
		\caption{Wave-packet reflection and accumulated local dissipation at the threshold coupling $\Delta=\Delta_c=\sqrt{5}/6$. (a)~Real and imaginary parts of $\psi_{x,A}$ at $t=27$, before the first boundary encounter, and (b)~at $t=43$, after reflection; black arrows indicate the propagation direction. (c)~Accumulated onsite dissipation $P_{x,\rm onsite}$ and (d)~accumulated hopping dissipation $P_{x,\rm hop}$, both evaluated as $t_f\to\infty$. Near either boundary, both contributions exhibit a period-2 high-low fringe pattern. Common parameters: $t_R=1$, $t_L=1/2$, $L=101$, $x_0=51$.}\label{figa3}
    \end{figure}

At $\Delta=\Delta_c$, the reflected packets return to the bulk decay envelope to leading order, yet their total norm temporarily falls below this envelope during the encounter [Fig.~2(a) in the main text]. Norm-preserving reflection fixes the net boundary response without requiring loss and gain to balance at every instant. Figure~\ref{figa3} shows the carrier structure before and after reflection, together with the accumulated near-edge dissipation.

Figures~\ref{figa3}(a,b) show that reflection increases the left packet's amplitude on chain $A$ and decreases the right packet's amplitude on that chain. The dominant propagating channels lie near $k=\pm\pi/2$, with group velocities $\mp(t_R+t_L)$ [Sec.~\ref{sec:bulkdecay}]. Reflection switches these channels, which carry different weights on chains $A$ and $B$, redistributing each packet's weight between the chains. The rapid oscillations of the real and imaginary parts show the carrier phases. During the encounter, the incident and reflected carriers overlap, producing density interference.

We focus on the left boundary; the right boundary follows analogously. First consider the propagating components outside the skin tail. Their normalized spinors $\phi_L$ and $\phi_R$ specify the amplitudes on chains $A$ and $B$. In the local overlap region, we approximate their slowly varying envelopes by a common $A(t)$. For $s=A,B$, the propagating contribution takes the form:

\begin{equation}
    \begin{aligned}
    \psi_{x,s}&\simeq A(t)\left[(\phi_L)_s e^{ix\pi/2}-(\phi_R)_s e^{-ix\pi/2}\right],\\
    |\psi_{x,s}|^2&\simeq |A(t)|^2\left[| (\phi_L)_s|^2+|(\phi_R)_s|^2
       -2(-1)^x\mathrm{Re}\big((\phi_L)_s(\phi_R)_s^*\big)\right].
    \end{aligned}
    \label{sintf}
    \end{equation}

The minus sign in the first line follows from the reflection amplitude $r_L=-\tan\theta=-1$ at $\Delta_c$ [Sec.~\ref{sec:boundscatt}]. The momentum difference $\pi$ makes the density interference alternate between neighboring sites. On either chain, the incident and reflected amplitudes are unequal, so the fringe minima remain nonzero.

The interference fringes in Eq.~(\ref{sintf}) are opposite on the two chains: a maximum on one coincides with a minimum on the other. Their fringe contributions therefore cancel in the total density $|\psi_{x,A}|^2+|\psi_{x,B}|^2$, since $\phi_R^\dagger\phi_L=0$.

Near the boundary, the orthogonal propagating components cannot satisfy the open-boundary condition alone, so the skin component introduced in Sec.~\ref{sec:boundscatt} is also required. Its spinor $\phi_1$ has a nonzero overlap with the incident spinor $\phi_L$, so their interference survives the sum over the two chains. The incident and skin components have spatial factors $i^x$ and $(-i)^x\lambda_c^{-x}$, respectively, giving their interference a factor $(-1)^x\lambda_c^{-x}$. This factor produces alternating high and low densities on neighboring sites, with a contrast that decays away from the boundary. Since the onsite loss rate is $2V$ times the total density [Eq.~(11) in the End Matter], its time integral gives the boundary fringes in Fig.~\ref{figa3}(c).

The hopping contribution depends on the product of neighboring amplitude magnitudes and the sine of their phase difference [Eq.~(12) in the End Matter]. Including the skin component in these products introduces terms proportional to $(-1)^x\lambda_c^{-x}$, giving the boundary fringes in Fig.~\ref{figa3}(d). For the parameters shown, the positive onsite contribution records loss, while the negative hopping contribution records gain.

During the encounter, the time-dependent balance between onsite loss and hopping-induced gain governs the instantaneous norm change, $\dot{\mathcal N}=-\sum_x\dot P_x$. The transient dip reflects a temporary departure from the bulk decay envelope. Once the packets leave the boundary region, they again follow the bulk decay envelope to leading order, with a vanishing boundary step $\delta h\simeq0$ [Eq.~(\ref{appStepReflection})].

\subsection{Overall dissipation behavior of NHEA}
\begin{figure}[ht]
		\centering
		\includegraphics[width=0.95\textwidth]{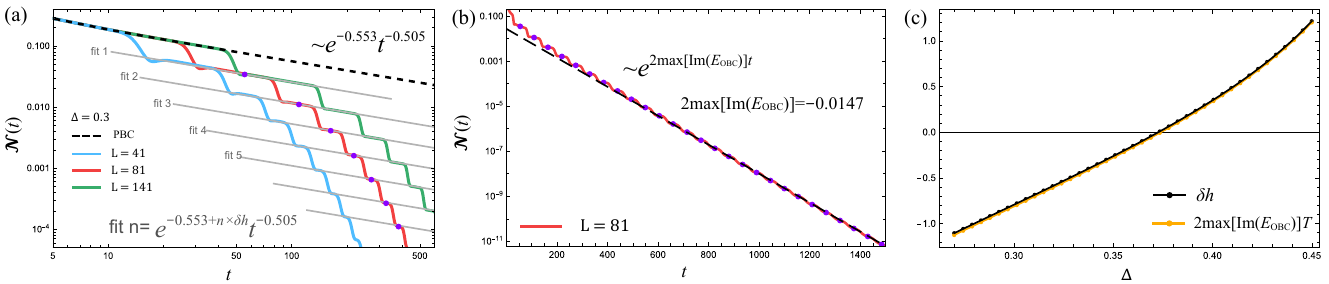}\\
		\caption{(a) Total norm $\mathcal{N}(t)$ versus time under PBCs (dashed) and OBCs (solid). The violet points mark samples at times close to the packets' returns to the midpoint (for $L=81$). (b) Longer-time OBC evolution for $L=81$ on a semilogarithmic scale; the dashed line shows the exponential envelope set by $2\max[\mathrm{Im}(E_{\rm OBC})]$. Panels (a,b) use $\Delta=0.3$. (c) Comparison of $\delta h$ and $2T\max[\mathrm{Im}(E_{\rm OBC})]$ as functions of $\Delta$. Panel (c) uses $L=81$. In all panels, $t_R=1$, $t_L=1/2$, and the initial state is $|x_0,B\rangle$ with $x_0=(L+1)/2$.}\label{figa6}
    \end{figure}

    The bulk decay envelope and the norm multiplier at each boundary encounter together produce the staircase dynamics. This description applies for fixed $0<\Delta<\Delta_{\rm PT}^{\rm PBC}$ with $L\ln\lambda\gg1$, while successive boundary encounters remain well separated. By examining the total norm $\mathcal{N}(t)$ as a function of time for different system sizes, we confirm that the $n$th decay segment between boundary encounters follows $e^{n\delta h}t^{-0.5}$, as shown in Fig.~\ref{figa6}(a). Here $\delta h$ is the increment between adjacent plateaus of $h(t)=\ln[\mathcal N_{\rm OBC}(t)/\mathcal N_{\rm PBC}(t)]$, set by the reflection factor in Eq.~(\ref{appStepReflection}). Each paired boundary encounter therefore multiplies the norm, relative to the bulk envelope, by $e^{\delta h}$, giving the factor $e^{n\delta h}$ after $n$ such encounters. The multiplier is independent of $L$ to leading order, whereas the interval between encounters increases with $L$.

    In Fig.~\ref{figa6}(a), the violet points sample the norm at times close to the packets' returns to the midpoint. During the resolved-packet stage, these samples obey $\mathcal N_{\rm OBC}(nT)\sim e^{n\delta h}(nT)^{-1/2}$, with $T=(L+1)/(t_R+t_L)$. The plotted $\Delta=0.3<\Delta_c$ illustrates negative boundary steps, while positive steps produce NHEA in the boundary-gain-dominated regime.

    At later times in a fixed finite chain, the eigenvalues with the largest imaginary part determine the exponential norm envelope, $\mathcal N_{\rm OBC}(t)\propto e^{2\max[\mathrm{Im}(E_{\rm OBC})]t}$, as illustrated by the dashed fit in Fig.~\ref{figa6}(b). Boundary scattering and finite-chain quantization connect this spectral rate to the earlier boundary steps through $\delta h\simeq2T\max[\mathrm{Im}(E_{\rm OBC})]$ [Eq.~(\ref{appStepSpectral})]. The factor of two converts the amplitude growth rate into the norm growth rate, and $T$ is the interval between successive paired boundary encounters. Figure~\ref{figa6}(c) compares these two quantities directly. A nonzero leading boundary step is independent of $L$, while the encounter interval grows as $T\propto L$, yielding an $L^{-1}$ spectral rate at fixed coupling. At $\Delta_c$, the leading step vanishes, leaving the finite-size residual loss derived in Sec.~\ref{sec:threshold}.

\section{Large-\texorpdfstring{$L$}{L} estimate of the number of complex OBC eigenenergies at \texorpdfstring{$\Delta=\Delta^\text{PBC}_\text{PT}$}{Delta = Delta(PT, PBC)}}\label{sec:complexnum}

At $\Delta=\Delta_{\rm PT}^{\rm PBC}\equiv t_R-t_L$, the PBC spectrum is purely real, while the open boundaries support complex-conjugate eigenvalues. In this section, we quantify their number through a large-$L$ phase-averaged estimate at fixed $0<t_L/t_R<1$. The strategy is to count the real OBC eigenvalues in the PBC band region $[-2t_R,2t_R]$ and subtract this number from the total $2L$.

At this coupling, the characteristic equation (\ref{sHbeta}) for the non-Bloch modes in Eq.~(\ref{sgbz}) gives
\begin{equation}
	\beta_{1,4}=\frac{E\pm\sqrt{E^2-4t_L^2}}{2t_L},\quad
	\beta_{2,3}=\frac{E\pm\sqrt{E^2-4t_R^2}}{2t_R},
	\label{beta1234}
\end{equation}
satisfying $\beta_1\beta_4=\beta_2\beta_3=1$. The PBC bands span $[-2t_R,2t_R]$, with an overlap window $[-2t_L,2t_L]$ and two non-overlap windows. These windows contain the full real OBC spectrum, since at $V=0$ the Hermitian part of the finite-chain Hamiltonian gives $|\operatorname{Re}E|\leq(t_R+t_L)\cos[\pi/(L+1)]+\Delta<2t_R$.

\begin{figure}[t]
	\centering
	\includegraphics[width=0.6\textwidth]{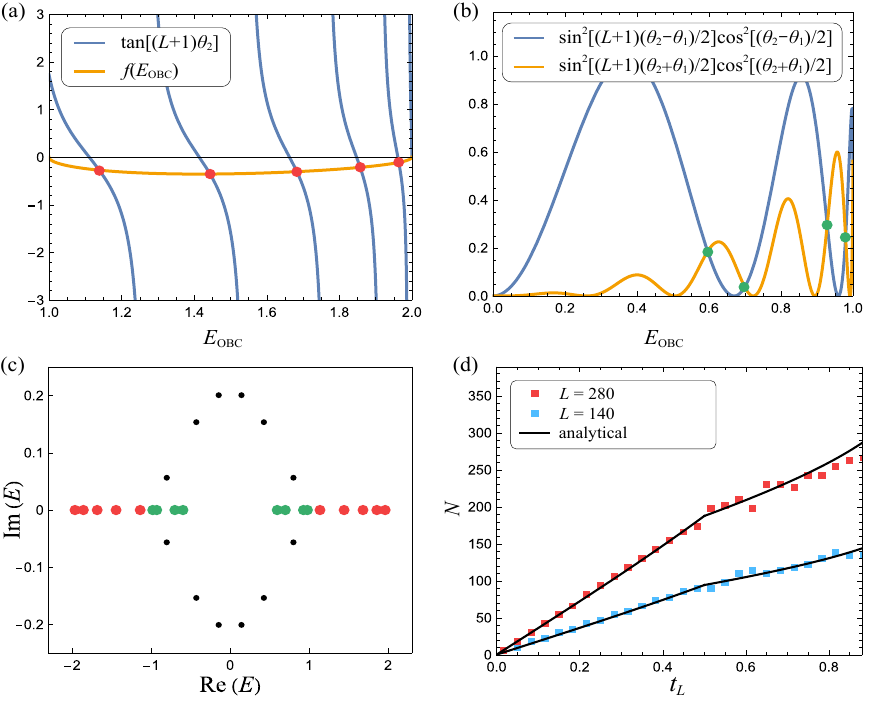}\\
	\caption{(a) Solutions of Eq.~(\ref{th2}) on $E_{\rm OBC}\in(2t_L,2t_R)$ (red points), where the orange curve is the negative function $f(E_{\rm OBC})$ defined in Eq.~(\ref{th2}), and (b) solutions of Eq.~(\ref{th1}) on $E_{\rm OBC}\in(0,2t_L)$ (green points), at $t_R=1$, $t_L=1/2$, $\Delta=1/2$, and $L=15$. The leading-order estimate gives $(L+1)\arccos(t_L/t_R)/\pi\approx 5.3$ roots in each positive-energy window; finite-size rounding gives five red and four green intersections in panels (a) and (b). (c) Corresponding OBC eigenvalues, with the red and green markers identifying the real eigenvalues counted in (a) and (b). (d) Number of complex OBC eigenvalues $N$ versus $t_L$ at $t_R=1$ and $\Delta=t_R-t_L$; the black curve is the compact analytical interpolation in Eq.~(\ref{nL}). The slope change at $t_L=1/2$ reflects the algebraic crossover of the piecewise approximation in Eq.~(\ref{cAapprox}), while the underlying OBC spectrum evolves smoothly without qualitative topological restructuring. }
	\label{figa4}
\end{figure}

First consider $E\in(2t_L,2t_R)$. Choose $\beta_1=1/\beta_4>1$ and $\beta_2=1/\beta_3=e^{i\theta_2}$, where $\theta_2=\arccos[E/(2t_R)]$. Keeping the leading-order $\beta_1^{L+1}$ terms in Eq.~(\ref{sobc}) gives
\begin{equation}
	X_{1,3}X_{2,4}\exp[-i(L+1)\theta_2]-X_{1,2}X_{3,4}\exp[i(L+1)\theta_2]\approx0.
	\label{th21}
\end{equation}
The coefficients $X_{1,3}X_{2,4}$ and $X_{1,2}X_{3,4}$ are complex conjugates in this window. Using their definitions in Eq.~(\ref{sobc}), we obtain the equivalent real phase condition
\begin{equation}
	\tan[(L+1)\theta_2]\approx f(E),\qquad
	f(E)=-\frac{\sqrt{E^2-4t_L^2}\sqrt{4t_R^2-E^2}}{E^2+4t_Lt_R}.
	\label{th2}
\end{equation}
In the window interior, the propagation phase $(L+1)\theta_2$ has an $O(L)$ slope with respect to $\theta_2$, whereas the boundary phase $\arctan f(E)$ has an $O(1)$ slope. Thus one root occurs per $\pi$ advance of the quantization phase $(L+1)\theta_2-\arctan f(E)$ to leading order. The positive non-overlap window therefore contains
$N_{\rm non}^{(+)}\simeq (L+1)\arccos(t_L/t_R)/\pi$ real roots, and the $E\to-E$ symmetry doubles this contribution [red points, Figs.~\ref{figa4}(a),(c)].

In the overlap region $E\in(0,2t_L)$, all four roots are propagating:
$\beta_1=1/\beta_4=e^{i\theta_1}$ and $\beta_2=1/\beta_3=e^{i\theta_2}$, with $\theta_1\in(0,\pi/2)$ and $\theta_2\in(\arccos(t_L/t_R),\pi/2)$. Eq.~(\ref{sobc}) reduces to
\begin{equation}
	\sin^{2}[(L+1)(\theta_2-\theta_1)/2]\cos^{2}[(\theta_2-\theta_1)/2]
	=
	\sin^{2}[(L+1)(\theta_2+\theta_1)/2]\cos^{2}[(\theta_2+\theta_1)/2],
	\label{th1}
\end{equation}
where $\theta_1=\arccos[E/(2t_L)]$ and $\theta_2=\arccos[(t_L/t_R)\cos\theta_1]$. To count the allowed standing waves, introduce
\begin{equation}
	U_2=\frac{(L+1)\theta_2}{2},\qquad U_1=\frac{(L+1)\theta_1}{2},\qquad
	\kappa=\tan\frac{\theta_1}{2}\tan\frac{\theta_2}{2}.
	\label{UVk}
\end{equation}
Since both cosine factors in Eq.~(\ref{th1}) are strictly positive on the open window, taking the square root of both sides with both signs and expanding the sine functions gives the two branches,
\begin{equation}
	\begin{gathered}
		\sin U_2\cos U_1-\kappa\cos U_2\sin U_1=0\qquad\text{or}\qquad \sin U_1\cos U_2-\kappa\cos U_1\sin U_2=0.
	\end{gathered}
	\label{branches}
\end{equation}
Here $\delta=\theta_2-\theta_1$ and $\Sigma=\theta_2+\theta_1$ are the difference and sum phases describing interference between the two propagation channels under common boundary matching. The reduction uses the identity $\cos(\Sigma/2)/\cos(\delta/2)=(1-\kappa)/(1+\kappa)$.

We count the roots of Eq.~(\ref{branches}) as $\theta_1$ runs from $0$ to $\pi/2$. Away from tangent poles, the first branch gives $U_2=\arctan(\kappa\tan U_1)+n\pi$. To follow this phase continuously, define $a_\kappa(u)$ as the continuous unwrapping of $\arctan(\kappa\tan u)$, with $a_\kappa(0)=0$ and $a_\kappa(u+\pi)=a_\kappa(u)+\pi$. The branch condition then reads $U_2=a_\kappa(U_1)+n\pi$, so a root occurs whenever the phase difference $\Phi\equiv U_2-a_\kappa(U_1)$ crosses an integer multiple of $\pi$. Differentiating the continuous phase $\Phi$ with respect to $\theta_1$ and retaining the leading-order terms in $L$ gives
\begin{equation}
	\Phi'\simeq\tfrac{1}{2}(L+1)\big[q-h_\kappa(U_1)\big],\qquad
	q\equiv\frac{d\theta_2}{d\theta_1}.
	\label{Phip}
\end{equation}
Here $h_\kappa(\tau)=\partial a_\kappa(\tau)/\partial\tau$ at fixed $\kappa$, where $\tau$ denotes the phase argument of $a_\kappa$; the contribution of $d\kappa/d\theta_1$ to $\Phi'$ is $O(1)$ in the window interior. Differentiating $E=2t_L\cos\theta_1=2t_R\cos\theta_2$ and $a_\kappa(\tau)$ gives the rates $q$ and $h_\kappa$ explicitly,
\begin{equation}
	q(\theta_1)=\frac{t_L\sin\theta_1}{t_R\sin\theta_2}=\frac{v_1}{v_2}\in(0,1),\qquad
	h_\kappa(\tau)=\frac{\kappa}{\cos^2\tau+\kappa^2\sin^2\tau} ,
	\label{qh}
\end{equation}
Physically, $q=v_1/v_2$ is the ratio of the two channels' group velocities $v_j=dE/d\theta_j$. The phase rate $h_\kappa(\tau)$ is $\pi$-periodic, ranges from $\kappa$ at $\tau=0$ to $1/\kappa$ at $\tau=\pi/2$, and has unit mean, $\langle h_\kappa\rangle\equiv\tfrac1\pi\int_0^\pi h_\kappa\,d\tau=1$.

The allowed phase values $\Phi=n\pi$ are separated by $\pi$. For the large-$L$ estimate, we replace these discrete values by their mean density $1/\pi$ in $\Phi$. A phase change $|d\Phi|$ then contributes $|d\Phi|/\pi$ roots on average, giving $N_1\simeq\tfrac1\pi\int_0^{\pi/2}|\Phi'|\,d\theta_1$ for the first branch. The absolute value counts both directions of phase crossing in this averaged description.

When $\theta_1$ increases by $2\pi/(L+1)$, the fast phase $U_1=(L+1)\theta_1/2$ advances by one period, while $q$ and $\kappa$ vary on the slow scale. Averaging the root density $|\Phi'|/\pi$ over this period gives $\tfrac{L+1}{2\pi}\langle|q-h_\kappa|\rangle$ roots per unit $\theta_1$. The second branch of Eq.~(\ref{branches}) gives $U_1=a_\kappa(U_2)+n\pi$, so its counting phase is $U_1-a_\kappa(U_2)$. Its leading derivative is $\tfrac{L+1}{2}[1-qh_\kappa(U_2)]$, giving $\tfrac{L+1}{2\pi}\langle|1-qh_\kappa|\rangle$. Adding the two root densities and integrating over $\theta_1\in(0,\pi/2)$ gives the estimated number $N_{\rm ov}^{(+)}$ of real OBC eigenvalues in the positive overlap window $E\in(0,2t_L)$,
\begin{equation}
	\begin{aligned}
		N_{\rm ov}^{(+)}&\simeq \frac{L+1}{\pi}\int_0^{\pi/2}\!A(q(\theta_1),\kappa(\theta_1))\,d\theta_1,\\
		A(q,\kappa)&=\frac{1}{2\pi}\int_0^\pi\!\big[\,|q-h_\kappa(\tau)|+|1-q\,h_\kappa(\tau)|\,\big]\,d\tau .
	\end{aligned}
	\label{Aint}
\end{equation}
The counting kernel $A(q,\kappa)$ is half the sum of the two period-averaged factors. Its slow arguments are $q=q(\theta_1)$ and $\kappa=\kappa(\theta_1)$ from Eqs.~(\ref{UVk}) and (\ref{qh}), with $\theta_2=\arccos[(t_L/t_R)\cos\theta_1]$; $\tau$ is the fast phase being averaged. The $E\to-E$ symmetry doubles the positive-energy contribution [green points, Figs.~\ref{figa4}(b),(c)].

Combining the positive and negative windows therefore gives $N_{\rm real}\simeq2(N_{\rm ov}^{(+)}+N_{\rm non}^{(+)})$. Subtracting $N_{\rm real}$ from the total $2L$ and retaining the leading term in $L$ gives
\begin{equation}
	N\simeq 2L\Big[\,1-\frac{1}{\pi}\arccos\frac{t_L}{t_R}-c_A\,\Big],\qquad
	c_A=\frac{1}{\pi}\int_0^{\pi/2}\!A(q(\theta_1),\kappa(\theta_1))\,d\theta_1 ,
	\label{nLgen}
\end{equation}
for the number of complex eigenvalues, counting both members of each conjugate pair. The bracketed coefficient $1-\arccos(t_L/t_R)/\pi-c_A$ therefore estimates the complex-eigenvalue fraction $N/(2L)$. The Hermitian endpoint $t_L=t_R$, $\Delta=0$ has $N=0$ and lies outside this fixed-$t_L/t_R<1$ large-$L$ approximation.

To make Eq.~(\ref{nLgen}) explicit we evaluate $c_A$. The counting kernel $A(q,\kappa)$ simplifies whenever the expressions $q-h_\kappa(\tau)$ and $1-qh_\kappa(\tau)$ inside the absolute values in Eq.~(\ref{Aint}) keep fixed signs over the $\tau$-integral. Because $h_\kappa$ ranges over $[\kappa,1/\kappa]$, the single condition $q\le\kappa$ forces $q\le h_\kappa$ and $q\,h_\kappa\le1$ for every $\tau$. The absolute values can then be evaluated explicitly, and using $\langle h_\kappa\rangle=1$,
\begin{equation}
	A=\frac{1}{2\pi}\int_0^\pi\!\big[(h_\kappa-q)+(1-q\,h_\kappa)\big]\,d\tau=1-q .
	\label{Aweak}
\end{equation}
From Eqs.~(\ref{UVk}) and (\ref{qh}), $q/\kappa=t_L(1+\cos\theta_1)/(t_R-t_L\cos\theta_1)$. This ratio decreases with $\theta_1$ and approaches $2t_L/(t_R-t_L)$ at the band edge $\theta_1\to0$. Thus $q\le\kappa$ throughout the overlap window when $2t_L/(t_R-t_L)\le1$, or equivalently $t_L/t_R\le1/3$. In this regime, $A=1-q$, and using $\int_0^{\pi/2}q\,d\theta_1=\theta_2(\pi/2)-\theta_2(0)=\pi/2-\arccos(t_L/t_R)$ gives
\begin{equation}
	c_A=\frac1\pi\int_0^{\pi/2}(1-q)\,d\theta_1=\frac1\pi\arccos\frac{t_L}{t_R}\qquad(t_L/t_R\le\tfrac13).
	\label{cAweak}
\end{equation}
Substitution into Eq.~(\ref{nLgen}) gives the small-overlap result $N\simeq4L\arcsin(t_L/t_R)/\pi$ for $t_L/t_R\leq1/3$.

For $t_L/t_R>1/3$ a band-edge region with $q>\kappa$ appears. There $q-h_\kappa(\tau)$ and $1-qh_\kappa(\tau)$ change sign during a $\tau$ period, so the two counting phases can reverse direction locally. The absolute values in Eq.~(\ref{Aint}) include these reversals in the averaged root count, yielding $A>1-q$ and raising $c_A$ above $\arccos(t_L/t_R)/\pi$. Relative to the $A=1-q$ baseline, the additional number $\Delta_{\rm ov}^{(+)}$ of real levels in the positive overlap window is estimated by
\begin{equation}
	\Delta_{\rm ov}^{(+)}=\frac{L+1}{\pi}\int_0^{\pi/2}\!\big[A(q(\theta_1),\kappa(\theta_1))-(1-q(\theta_1))\big]\,d\theta_1\geq0.
	\label{dcorr}
\end{equation}
These additional real levels lower the complex-eigenvalue estimate in Eq.~(\ref{nLgen}) below the continuation of the small-overlap result $4L\arcsin(t_L/t_R)/\pi$.

Beyond this onset, the correction in Eq.~(\ref{dcorr}) counteracts the decrease of the small-overlap expression Eq.~(\ref{cAweak}), producing a shallow numerical minimum of $c_A$ near $1/3$. For a coarse description of $1/3<t_L/t_R<1$, we approximate the corrected coefficient by this near-minimum value, $c_A\simeq1/3$. Near the lower end of this interval, the constant underestimates the integral value of $c_A$. We therefore use the continuation of the small-overlap expression to connect to the constant approximation through the crossover interval $1/3<t_L/t_R<1/2$, adopting the continuous empirical interpolation
\begin{equation}
	c_A\simeq\max\!\Big[\frac1\pi\arccos\frac{t_L}{t_R},\ \frac13\Big]
	\label{cAapprox}
\end{equation}
which reproduces Eq.~(\ref{cAweak}) for $t_L/t_R\le1/3$. The two interpolation branches meet when $\arccos(t_L/t_R)/\pi=1/3$, giving $t_L/t_R=1/2$. This value is set by the chosen constant approximation; it has no additional dependence on $L$ or the overall hopping scale. The exact onset of the nonzero integral correction in Eq.~(\ref{dcorr}) remains $t_L/t_R=1/3$. Substituting Eq.~(\ref{cAapprox}) into Eq.~(\ref{nLgen}) gives
\begin{equation}
	N\simeq
	\begin{cases}
		\dfrac{4L}{\pi}\arcsin\dfrac{t_L}{t_R}, & t_L/t_R\le\tfrac12 ,\\[2.2mm]
		2L\Big(\dfrac23-\dfrac1\pi\arccos\dfrac{t_L}{t_R}\Big), & t_L/t_R\ge\tfrac12 ,
	\end{cases}
	\label{nL}
\end{equation}
Eq.~(\ref{nL}) provides a compact analytical estimate of the size and hopping-ratio dependence of $N$, compared with finite-chain counts in Fig.~\ref{figa4}(d).

The mechanism is band competition in the overlap window. Two real PBC bands share $[-2t_L,2t_L]$, but their standing waves must satisfy a common four-root interference condition at the open boundaries. The resulting deficit of real levels is carried by complex-conjugate pairs, since the finite-chain Hamiltonian is real at $V=0$. At fixed $0<t_L/t_R<1$, Eq.~(\ref{nLgen}) estimates an extensive complex sector even though the PBC spectrum is entirely real, quantifying the fraction of modes affected by boundary-induced complexification.

\section{Complex OBC spectrum and real PBC spectrum in a coupled SSH model}

    Here we provide a compact four-band test showing that the real-PBC/complex-OBC spectral reconstruction extends beyond the coupled HN model. This example tests the edge-induced $\mathcal{PT}$-breaking mechanism described in the main text, rather than establishing NHEA wave-packet dynamics in the Su--Schrieffer--Heeger (SSH) model. We couple two non-Hermitian SSH chains with opposite intracell nonreciprocity through the equal-sublattice hopping $\Delta$. In the basis $(1A,1B,2A,2B)$, the Bloch Hamiltonian is
    \begin{equation}
		H(k)=\left( \begin{array}{cccc}
		0 & t_1+\gamma+t_2e^{-ik}& \Delta& 0 \\
		t_1-\gamma+t_2e^{ik}& 0& 0& \Delta \\
            \Delta & 0& 0& t_1-\gamma+t_2e^{-ik} \\
            0 & \Delta& t_1+\gamma+t_2e^{ik}& 0
		\end{array}\right).
		\label{hkssh}
	\end{equation}

Let $\tau_i$ and $\sigma_i$ act on the chain and sublattice spaces, respectively. At $\Delta=0$, the two SSH chains have generalized Brillouin-zone radii $\rho_1=\sqrt{(t_1-\gamma)/(t_1+\gamma)}$ and $\rho_2=\rho_1^{-1}$, demonstrating their opposite skin-accumulation directions. In real space, parity combines $x\mapsto L+1-x$ with the internal transformation $U_{\mathcal P}=\tau_x\otimes\sigma_x$, while $\mathcal T=K$. For real parameters,

    \[
    U_{\mathcal P}H^*(k)U_{\mathcal P}^{-1}=H(k),
    \]
    so both the Bloch and finite-chain Hamiltonians possess exact $\mathcal{PT}$ symmetry. The coupling $\Delta$ connects equal sublattices and places the two oppositely nonreciprocal channels under the same open boundaries.

    To isolate the bulk-overlap mechanism from ordinary SSH end states, we use $t_1=5/4$, $t_2=1$, and $\gamma=\Delta=1/10$. The isolated chains are topologically trivial, since $t_2<\sqrt{t_1^2-\gamma^2}$, and all four PBC bands are real and nondegenerate. The two positive-energy bands span $[1/\sqrt{80},\sqrt{369/80}]$ and $[3/\sqrt{80},21/\sqrt{80}]$, respectively. Their overlap is therefore

    \[
    I_{\rm ov}^{+}=[3/\sqrt{80},\sqrt{369/80}]
    \simeq[0.3354,2.1477],
    \qquad
    I_{\rm ov}^{-}=-I_{\rm ov}^{+}.
    \]

    For a real energy in either overlap window, each PBC band contributes a propagating pair $\beta=e^{\pm ik}$, so all four roots of $\det[H(\beta)-E]=0$ have unit modulus. For example, at $E=1$ all four roots satisfy $|\beta_j|=1$, whereas at $E=0.2$, which belongs to only one band, their moduli are approximately $\{0.7936,1,1,1.2601\}$. Thus the overlap removes the exponential root hierarchy and requires the full four-wave open-boundary matching. The overlap supplies this root structure but is not sufficient by itself: opposite nonreciprocity, finite interchain coupling, and common open boundaries remain essential.

    For $L=20$ unit cells, the OBC Hamiltonian has dimension $4L=80$ and contains $N_{\rm complex}=36$ complex eigenvalues, forming 18 complex-conjugate pairs whose real parts all lie in $I_{\rm ov}^{+}\cup I_{\rm ov}^{-}$. The conjugate pairing follows because the finite-chain Hamiltonian is real. As a finite-size overlap check, at fixed $t_1=5/4$, $t_2=1$, $\gamma=1/10$, and $L=20$, the positive-energy overlap width decreases through $1.853$, $1.812$, $1.771$, and $1.731$ for $\Delta=0.08$, $0.10$, $0.12$, and $0.14$, while $N_{\rm complex}=52$, $36$, $24$, and $4$, respectively. All four PBC spectra remain real along this parameter path. This comparison demonstrates the multiband spectral mechanism without claiming that overlap alone, or every finite $L$, necessarily produces complex eigenvalues.

    \begin{figure}[ht]
		\centering
		\includegraphics[width=0.8\textwidth]{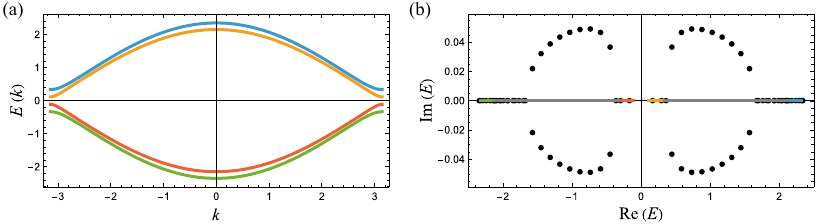}\\
		\caption{(a) Four entirely real PBC bands of the coupled SSH model; orange/blue and red/green denote the inner and outer positive/negative-energy branches, respectively. (b) OBC eigenvalues (black dots) for $L=20$ unit cells, corresponding to a $4L=80$ dimensional matrix; the colored segments on $\operatorname{Im}E=0$ mark the nonoverlapping parts of the PBC bands, and the thick gray segments mark the overlap $I_{\rm ov}^{+}\cup I_{\rm ov}^{-}$. Eighteen complex-conjugate OBC pairs occur, all with real parts inside the PBC-band overlap. The isolated SSH chains are topologically trivial, $t_2<\sqrt{t_1^2-\gamma^2}$, so the complexification is not attributable to ordinary SSH end modes and instead provides an independent test of the four-root boundary mechanism. Here $t_1=5/4$, $t_2=1$, $\gamma=\Delta=1/10$.}\label{figssh}
    \end{figure}

\bibliography{NHEA_References}